\documentclass{ptptex}

\usepackage{amsmath,amscd,amssymb,graphicx}
\usepackage{epsfig}

\makeatletter
\@addtoreset{equation}{section}

\makeatother

\title{Vacua of Massive Hyper-K\"ahler Sigma Models 
with Non-Abelian Quotient}
\author{Masato Arai$^{1,}$\footnote{Present address: Department of
Physics, Tokyo Institute of Technology, Tokyo 152-8551, Japan.}, 
Muneto Nitta$^{2,*}$ and Norisuke Sakai$^{3}$}
\inst{$^{1}$Institute of Physics, AS CR, 
  182 21, Praha 8, Czech Republic \\
$^{2}$Department of Physics, Purdue University, 
West Lafayette, IN 47907-1396, USA\\
$^{3}$Department of Physics, Tokyo Institute of 
Technology,
Tokyo 152-8551, Japan}
\abst{
The Higgs branch of ${\cal N}=2$ supersymmetric gauge theories 
with non-Abelian gauge groups 
are described by hyper-K\"ahler (HK) nonlinear sigma models 
with potential terms. 
Using the non-Abelian HK quotient with respect to 
$U(M)$ and $SU(M)$ gauge groups, we derive 
the massive HK sigma models that are not toric in the ${\cal N}=1$
 superfield formalism and the harmonic superspace formalism.
The $U(M)$ quotient gives $N!/[M! (N-M)!]$ discrete vacua 
that may allow various types of domain walls, 
whereas the $SU(M)$ quotient gives no discrete vacua. 
}

\begin{document}
\maketitle
\newcommand {\beq}{\begin{eqnarray}}
\newcommand {\eeq}{\end{eqnarray}}
\newcommand {\non}{\nonumber\\}
\newcommand {\eq}[1]{\label {eq.#1}}
\newcommand {\defeq}{\stackrel{\rm def}{=}}
\newcommand {\gto}{\stackrel{g}{\to}}
\newcommand {\hto}{\stackrel{h}{\to}}
\newcommand {\1}[1]{\frac{1}{#1}}
\newcommand {\2}[1]{\frac{i}{#1}}
\newcommand {\thb}{\bar{\theta}}
\newcommand {\ps}{\psi}
\newcommand {\psb}{\bar{\psi}}

\newcommand {\ph}{\varphi}
\newcommand {\phs}[1]{\varphi^{*#1}}
\newcommand {\sig}{\sigma}
\newcommand {\sigb}{\bar{\sigma}}
\newcommand {\Ph}{\Phi}
\newcommand {\Phd}{\Phi^{\dagger}}
\newcommand {\Sig}{\Sigma}
\newcommand {\Phm}{{\mit\Phi}}
\newcommand {\eps}{\varepsilon}
\newcommand {\del}{\partial}
\newcommand {\dagg}{^{\dagger}}
\newcommand {\pri}{^{\prime}}
\newcommand {\prip}{^{\prime\prime}}
\newcommand {\pripp}{^{\prime\prime\prime}}
\newcommand {\prippp}{^{\prime\prime\prime\prime}}

\newcommand {\pripppp}{^{\prime\prime\prime\prime\prime}}
\newcommand {\delb}{\bar{\partial}}
\newcommand {\zb}{\bar{z}}
\newcommand {\mub}{\bar{\mu}}
\newcommand {\nub}{\bar{\nu}}
\newcommand {\lam}{\lambda}
\newcommand {\lamb}{\bar{\lambda}}
\newcommand {\kap}{\kappa}
\newcommand {\kapb}{\bar{\kappa}}
\newcommand {\xib}{\bar{\xi}}
\newcommand {\ep}{\epsilon}
\newcommand {\epb}{\bar{\epsilon}}
\newcommand {\Ga}{\Gamma}
\newcommand {\rhob}{\bar{\rho}}
\newcommand {\etab}{\bar{\eta}}
\newcommand {\chib}{\bar{\chi}}
\newcommand {\tht}{\tilde{\th}}
\newcommand {\zbasis}[1]{\del/\del z^{#1}}
\newcommand {\zbbasis}[1]{\del/\del \bar{z}^{#1}}
\newcommand {\vecv}{\vec{v}^{\, \prime}}
\newcommand {\vecvd}{\vec{v}^{\, \prime \dagger}}
\newcommand {\vecvs}{\vec{v}^{\, \prime *}}
\newcommand {\alpht}{\tilde{\alpha}}
\newcommand {\xipd}{\xi^{\prime\dagger}}
\newcommand {\pris}{^{\prime *}}
\newcommand {\prid}{^{\prime \dagger}}
\newcommand {\Jto}{\stackrel{J}{\to}}
\newcommand {\vprid}{v^{\prime 2}}
\newcommand {\vpriq}{v^{\prime 4}}
\newcommand {\vt}{\tilde{v}}
\newcommand {\vecvt}{\vec{\tilde{v}}}
\newcommand {\vecpht}{\vec{\tilde{\phi}}}
\newcommand {\pht}{\tilde{\phi}}
\newcommand {\goto}{\stackrel{g_0}{\to}}
\newcommand {\tr}{{\rm tr}\,}
\newcommand {\GC}{G^{\bf C}}
\newcommand {\HC}{H^{\bf C}}
\newcommand{\vs}[1]{\vspace{#1 mm}}
\newcommand{\hs}[1]{\hspace{#1 mm}}
\newcommand{\al}{\alpha}
\newcommand{\be}{\beta}
\newcommand{\Lam}{\Lambda}
\newcommand{\kahler}{K\"ahler }
\newcommand{\con}[1]{{\Gamma^{#1}}}
\newcommand {\dellr}{\stackrel{\leftrightarrow}{\partial}}


\section{Introduction}\label{INTRO}

Possibilities of large extra dimensions \cite{LED,RS} 
has renewed interest in field theories in 
higher-dimensional systems. 
In this brane-world scenario, our four-dimensional world 
is realized 
on an extended topological defect such as a wall (brane). 
Supersymmetry (SUSY) has been an extremely useful 
principle for building unified models beyond 
the standard model \cite{DGSW}. 
In SUSY theories, topological defects can often 
be obtained as BPS states \cite{BPS}, which preserve 
part of the 
SUSY \cite{WittenOlive}. 
The BPS states play an important role in exploring nonperturbative 
effects in SUSY gauge theories \cite{SW}. 
The Higgs branch of the SUSY gauge theories can often 
be described by SUSY nonlinear sigma models 
\cite{APS, AP}. 
In turn, these SUSY nonlinear sigma models are useful 
for obtaining topological defects, such as domain walls and 
flux tubes \cite{SY,EY}. 
Supersymmetric theories in systems of greater than four
dimensions require at least eight supercharges. 
The scalar and spinor matter fields can be described by means 
of hypermultiplets in theories with eight SUSY. 
Recently, we formulated ${1 \over 2}$ 
single-BPS domain walls in an eight SUSY model in four 
dimensions \cite{ANNS}. 
Moreover, we have also succeeded in constructing the 
${1 \over 2}$ BPS 
wall \cite{AFNS}
and BPS multi-walls \cite{EFNS}
consistently in five-dimensional supergravity. 
Before discussing the SUSY five-dimensional theories, 
it is useful to consider models with eight SUSY 
in four dimensions without gravity.

Theories with eight SUSY are so restrictive that 
the nontrivial interactions require the nonlinearity of 
the kinetic term (the nonlinear sigma model) 
if there are only hypermultiplets. 
Target manifolds of the SUSY nonlinear sigma models 
with eight SUSY must be hyper-K\"ahler (HK) 
manifolds~\cite{Zu,AF1}. 
To obtain a wall solution, we need a nontrivial 
 potential. 
The potential term for any HK sigma model is severely restricted 
by eight SUSY~\cite{AF2}. 
It can be written as the square of the tri-holomorphic 
Killing vector in the on-shell formulation, and 
it can be understood in terms of 
the Scherk-Schwarz reduction~\cite{SS} from 
six dimensions~\cite{ST}. 
These models are called ``massive HK nonlinear sigma models''.  
Contrastingly, in two dimensions, 
${\cal N}=2$ SUSY (four SUSY) sigma models 
can have a similar potential term~\cite{AF2}, 
in addition to the 
potential that can be derived from the superpotential 
and can be constructed off-shell 
in the superfield formalism~\cite{Ga}. 
Solitons in such models are discussed in Ref. \citen{LS}.
The off-shell formulation of ${\cal N}=2$ massive HK sigma models 
in four dimensions 
was given recently~\cite{ANNS}. 
An exact BPS solution of the wall junction was also constructed 
recently in an ${\cal N}=2$ theory with hypermultiplets and 
a vector multiplet \cite{KakimotoSakai}. 

A large class of the HK manifold is given by 
toric HK manifolds that are defined as HK manifolds 
of real dimension $4n$ 
admitting $n$ mutually commuting 
Abelian tri-holomorphic 
isometries~\cite{LR, PP, GGR, toric}.
In Ref.~\citen{GTT1}, 
the massive HK sigma model on any toric HK manifold 
was derived in the component formalism in four-dimensional 
spacetime. 
The solution of the $N$ parallel domain walls 
was found for $T^*{\bf C}P^{N-1}$, 
which is a toric HK manifold, 
and its moduli space was constructed \cite{GTT2}. 
For this massive $T^*{\bf C}P^{N-1}$ model, 
the dynamics of the walls were studied \cite{To} and 
the number of zero modes were calculated using 
the index theorem \cite{Le}.
The off-shell formulation of 
the massive $T^*{\bf C}P^{N-1}$ model in four-dimensional 
spacetime was obtained for both 
${\cal N}=1$ superfields and ${\cal N}=2$ superfields 
in harmonic superspace formalism (HSF) \cite{ANNS}. 
Other interesting solitons, like solutions representing
strings ending on walls, 
wall intersections and 
string intersections 
were also considered in the 
toric HK nonlinear sigma models 
\cite{GTT1,AT,GPTT,NNS1,PT}.

The potential term of 
the massive $T^*{\bf C}P^{N-1}$ model 
comes from the mass terms of the 
hypermultiplets when the nonlinear sigma model is 
constructed as the quotient with respect to the 
$U(1)$ gauge group~\cite{To,ANNS}. 
We call this formulation of massive HK sigma models 
 ``the massive HK quotient method,''
 since the massless case is identified to a HK quotient found 
 in Refs.~\citen{LR} and \citen{HKLR1}. 
One of the advantages of our massive HK quotient 
 is that the off-shell formulation of the SUSY 
 nonlinear sigma models is possible~\cite{ANNS}. 
The off-shell formulation is powerful for extending models
 to those with other isometries and/or gauge symmetries 
 and to those coupled to gravity,
 because in it (part of) SUSY is manifest.
Any {\sl toric} HK manifold can be constructed 
 using an {\sl Abelian} HK quotient~\cite{toric,GGR}.
Therefore, an off-shell formulation of general 
 massive toric HK sigma models~\cite{GTT1} 
 can be obtained 
 using the massive HK quotient with
Abelian gauge theories. 
 By contrast, 
 a HK nonlinear sigma model other than 
 the toric HK target manifolds has been 
 obtained by Lindstr\"om and Ro\v{c}ek~\cite{LR} 
 as a quotient using a non-Abelian gauge group 
 for the {\sl massless} 
 case only (without potential terms). 

The purpose of this paper is to 
 investigate {\sl massive} HK sigma models 
 in four-dimensional 
 spacetime using the quotient of the SUSY QCD 
 with respect to a
 {\sl non-Abelian} gauge group. 
The models are no longer toric HK manifolds, 
 and turn out to be 
 the cotangent bundle over 
 the complex Grassmann manifold $T^* G_{N,M}$ 
 and its generalization. 
The former model is the massive generalization of 
 the massless nonlinear sigma model 
 presented in Ref.~\citen{LR}. 
We obtain potential terms for this massive HK 
 nonlinear sigma model 
 and investigate the vacuum structure of 
 the massive $T^* G_{N,M}$ model in detail. 
We find that this model has $N!/[M!(N-M)!]$ discrete vacua 
 which are characterized by 
 mutually orthogonal $M$-dimensional complex hyperplanes 
 in ${\bf C}^N$.
We therefore can 
 expect rich structure for wall solutions in this model.
By considering the $SU(M)$ gauge group,  
 we obtain  a generalization of
the massive $T^* G_{N,M}$ model, 
the massive HK nonlinear sigma model 
with the quaternionic line bundle over the 
$T^* G_{N,M}$ as the target space. 
The vacua of this model are trivial, and hence 
we cannot expect interesting solitons.

There exists a similar model with the same number of vacua,  
namely the two-dimensional ${\cal N}=2$ 
SUSY (four SUSY) Grassmann $G_{N,M}$ sigma model 
with twisted mass~\cite{HH}. 
We believe that
 the target space of this model is obtained 
 by truncating our model on $T^*G_{N,M}$ to its base manifold, 
 so that they have the same number of vacua. 
If we find solitons in our model, 
 it may also be the case that
 they play interesting roles 
in this kind of two dimensional model.

The rest of this paper is organized as follows.
In $\S$2, we present the massive HK quotient with 
${\cal N}=1$ superfields. 
The vacuum structure of 
the massive $T^* G_{N,M}$ model is studied in 
$\S$3. 
The massive HK quotient is described in the Wess-Zumino gauge 
in $\S$4. 
Section 5 is devoted to a generalization of the massive 
HK quotient with respect to the $SU$ gauge group. 
Our model is derived in the harmonic superspace formalism 
 \cite{Ivanov,Ivanov2} in $\S$6. 
A summary and discussion are given in $\S$7.


\section{Massive HK quotient with respect to the 
 $U(M)$ gauge group}
We consider ${\cal N}=2$ SUSY QCD with 
$N$ flavors and a $U(M)$ gauge group.
In terms of ${\cal N}=1$ superfields, 
the $NM$ ${\cal N} =2$ hypermultiplets
 \footnote{In the following, we consider only 
 the case $N>M$. 
 This requirement is necessary in order for
 the nonlinear sigma models to be well-defined, 
 since dimensions of target
 space is given by $4(N-M)$.} 
can be decomposed 
into $(N \times M)$- and $(M \times N)$-matrix 
chiral superfields $\Phi(x,\theta,\thb)$ 
and $\Psi(x,\theta,\thb)$, and 
the ${\cal N}=2$ vector multiplets for the 
$U(M)$ gauge symmetry 
can be decomposed into $M\times M$ matrices of 
the ${\cal N}=1$ vector superfields 
$V=V^A(x,\theta,\thb) T_A$ 
and chiral superfields 
$\Sigma = \Sigma^A (x,\theta,\thb) T_A$,  
with the 
 $M\times M$ matrices $T_A~(A=1,\cdots,M^2)$ of the fundamental 
representation of the generators of the $U(M)$ gauge group.
The $U(M)$ gauge transformation is given by 
\beq
 &&e^V \to e^{V'} = 
   e^{-i \Lambda} e^V e^{i\Lambda\dagg},\hs{5}
   \Sigma \to \Sigma' = e^{-i \Lambda} \Sigma e^{i \Lambda}, 
   \label{gauge-tr-vector} \\
 &&\Phi \to \Phi' = \Phi e^{i \Lambda}, \hs{5}
 \Psi \to \Psi' = e^{-i \Lambda} \Psi, \hs{5} 
   \label{gauge-tr-hyper}
\eeq
where $\Lambda = \Lambda^A (x,\theta,\thb) T_A$ 
are the chiral superfields of the gauge parameters.  
Note that this gauge symmetry is actually complexified 
into $U(M)^{\bf C} = GL(M,{\bf C})$, 
because the scalar components of $\Lambda$ 
are complex fields.

Because we focus on the Higgs branch, 
we can take the strong coupling limit, $g \rightarrow \infty$, 
and throw away the kinetic term for gauge multiplets. 
Then, the Lagrangian is given by
\beq
&& {\cal L} = \int d^4 \theta
 \left[ \tr (\Phi\dagg\Phi e^V )  
 + \tr (\Psi\Psi\dagg e^{-V}) - c\, \tr V \right]  \non
&&\hs{5} + \left[ \int d^2\theta \,
       \left(\tr \left\{ \Sigma (\Psi \Phi - b {\bf 1}_M) \right\} 
      + \sum_{a=1}^{N-1} m_a \tr (\Psi H_a \Phi)\right) 
         + {\rm c.c.}\right] ,
\label{linear}
\eeq
where we have absorbed a common hypermultiplet mass 
into the field $\Sigma$, 
the quantities $m_a$ ($a=1,\cdots,N-1$) 
are complex mass parameters, 
and $H_a$ are diagonal traceless matrices,  
interpreted as the Cartan generators of $SU(N)$ below. 
The electric and magnetic Fayet-Iliopoulos (FI) parameters 
are denoted $c\in {\bf R}_{\geq 0}$ and $b \in {\bf C}$, respectively.  
In the limit $m_a\rightarrow 0$ for all $a$, 
the model has the global (flavor) $SU(N)$ symmetry 
expressed by
\beq
 \Phi \to \Phi' = g \Phi, \hs{5}
 \Psi \to \Psi' = \Psi g^{-1}, \hs{5} 
 g \in SU(N)\;, \label{global-sym}
\eeq
with $V$ and $\Sigma$ unchanged.\footnote{
When $b=0$, there exists an additional $U(1)$ given by 
$\Phi \to \Phi' = e^{i\alpha} \Phi, 
\Psi \to \Psi' = e^{i \alpha}\Psi,
\Sigma \to \Sigma' = e^{- 2 i \alpha}\Sigma$,  
but this is inconsistent 
with ${\cal N}=2$ SUSY.
}
For nonzero $m_a$, 
this $SU(N)$ symmetry is explicitly 
broken down to $U(1)^{N-1}$, generated by $H_a$.

Setting $\Psi=0$ in the Lagrangian (\ref{linear}),
it reduces to the \kahler quotient construction 
of the (complex) Grassmann manifold 
$G_{N,M} = SU(N)/[SU(M) \times SU(N-M) \times U(1)]$~\cite{HN}, 
and this suggests that the full Lagrangian is 
related to $G_{N,M}$.
Actually, for the massless case, this Lagrangian  
reduces to the HK quotient construction of 
$T^* G_{N,M}$~\cite{LR}, 
generalizing the $U(1)$ HK quotient construction~\cite{CF,AF3,RT} 
for $T^*{\bf C}P^{N-1}$ ($M=1$) 
with the Calabi metric~\cite{Ca}.

Since we have introduced the mass parameter through the $H_a$ 
generator, we will eventually obtain a potential term 
that is the square of the tri-holomorphic Killing vector 
corresponding to a linear combination of the matrices $H_a$, 
after eliminating the vector multiplet.

Next, we eliminate the auxiliary superfields 
$V$ and $\Sigma$ in the superfield formalism.
From Eq.~(\ref{linear}), we find that 
 their equations of motion read
\footnote{ 
Here we take $\delta X \equiv e^{-V} \delta e^V$ as 
an infinitesimal parameter of variation. 
Then, the equations 
$\delta \tr (\Phi\dagg\Phi e^V) = \tr(\Phi\dagg\Phi e^V \delta X)$ 
and 
$\delta \tr (\Psi\Psi\dagg e^{-V}) = 
- \tr(e^{-V}\Psi\Psi\dagg \delta X)$ hold. \label{var-V}
}
\begin{eqnarray}
 && {\partial {\cal L} \over \partial V} 
 = \Phi\dagg\Phi e^V - e^{-V} \Psi\Psi\dagg - c {\bf 1}_M = 0 \; , 
   \label{EOM-V}\\
 && {\partial {\cal L} \over \partial \Sigma} 
 = \Psi \Phi - b {\bf 1}_M = 0\;.
   \label{EOM-sig}
\end{eqnarray}
Multiplying Eq.~(\ref{EOM-V}) by 
 $\Phi^\dagger\Phi e^V$ from the left, we find
\begin{eqnarray}
 \left(\Phi^\dagger\Phi e^V - {c \over 2}{\bf 1}_M\right)^2
 -{c^2 \over 4}{\bf 1}_M-\Phi^\dagger\Phi\Psi\Psi^\dagger=0\,.
\end{eqnarray}
From the above equation, $V$ can be easily solved, and we have 
\beq
 e^V = {c\over 2} (\Phi\dagg\Phi)^{-1} 
  \left({\bf 1}_M  \pm \sqrt {{\bf 1}_M 
     + {4\over c^2} \Phi\dagg\Phi \Psi\Psi\dagg}\right) \,. 
  \label{sol-V}
\eeq 
Substituting this back into (\ref{linear}), 
we obtain the K\"ahler potential for 
the Lindstr\"{o}m-Ro\v{c}ek metric,~\cite{LR} 
\begin{equation}
 K = c\, \tr \sqrt{{\bf 1}_M + {4\over c^2} \Phi\dagg\Phi \Psi\Psi\dagg} 
   - c\, \tr \log \left( {\bf 1}_M 
    + \sqrt{{\bf 1}_M + {4\over c^2} \Phi\dagg\Phi \Psi\Psi\dagg}\right)
   + c\, \tr \log \Phi\dagg\Phi  \;, \label{kahler}
\end{equation}
where we have used the formula 
$\tr \log AB = \tr \log A + \tr \log B$, which holds 
for any square matrices $A$ and $B$,
and we have omitted constant terms because they vanish under 
the superspace integration.
Here we have chosen the plus sign in Eq.~(\ref{sol-V})  
 to realize the positivity of the metric. 
Using the $N\times N$ meson matrix of the gauge invariant,
\beq
 {\cal M} = \Phi\Psi \;,
\eeq
the \kahler potential can be rewritten as that 
in Ref. \citen{AP},
\beq
 &&K = c\, {M \over N} \tr_{N\times N} 
       \sqrt{{\bf 1}_N 
       + {N \over M}{4\over c^2} {\cal M} {\cal M}\dagg} 
   - c\, {M \over N}\tr_{N\times N} \log \left( {\bf 1}_N 
       + \sqrt{{\bf 1}_N 
       + {N \over M}{4\over c^2} 
       {\cal M}{\cal M}\dagg } \right) \non
  && \hs{5}
  + c\, \tr \log \Phi\dagg\Phi  \; , \label{kahler2}
\eeq
where we have made explicit that 
the traces are taken in $N \times N$ matrices, 
whereas the previous ones $M \times M$.

The \kahler potential (\ref{kahler}) or (\ref{kahler2}) 
is strictly invariant under the full global (flavor) 
$SU(N)$ symmetry (\ref{global-sym}), 
because the mass term in the Lagrangian 
(\ref{linear}) does not affect the D-term. 
In Eqs.~(\ref{kahler}) and (\ref{kahler2}), 
the first two terms are strictly invariant under the gauge 
transformation (\ref{gauge-tr-hyper}) for the matter fields, 
but the third term induces the \kahler transformation
\beq
 K \to K' = K + i c\, \tr \Lambda 
              - i c\, \tr \Lambda\dagg \;. 
\eeq

Fixing this complexified $U(M)$ gauge symmetry 
and solving the constraint (\ref{EOM-sig}), 
we obtain the Lagrangian of the nonlinear sigma model
in terms of independent superfields. 
For this purpose, we should consider the two cases 
$b = 0$ and $b \neq 0$ separately.
Note that the physics does not depend on this choice, 
 as the two cases are transformed into each other
 under the $SU(2)_R$ transformation.
Although this transformation does not preserve the holomorphy,
 these two cases merely involve
 different complex structures. 
\\
i) $b = 0$. In this case, the gauge can be fixed as
\beq
 \Phi = \begin{pmatrix}
           {\bf 1}_M \cr \ph
        \end{pmatrix}
         \;, \hs{5} 
 \Psi = (- \psi\ph, \psi) \;, \label{fixing1}
\eeq
with $\ph$ and $\psi$ being 
$[(N-M)\times M]$- and $[M\times (N-M)]$-matrix chiral 
superfields, respectively. 
Then, the superpotential becomes
\begin{eqnarray}
 &&W = \sum_a m_a \tr \left[
   (- \psi\ph, \psi) H_a \begin{pmatrix}
                           {\bf 1}_M \cr \ph
                         \end{pmatrix} 
  \right]  
  = \sum_a m_a \tr \left[ H_a
    \begin{pmatrix}
          -\psi\ph & \psi \cr 
          - \ph\psi\ph & \ph\psi
    \end{pmatrix}
      \right]\,. 
         \nonumber \\
 \label{superpot1}
\end{eqnarray}
ii) $b\neq 0$. 
In this case, we can take the gauge as~\cite{LR}
\beq
 \Phi = \begin{pmatrix}
          {\bf 1}_M \cr \ph
        \end{pmatrix}  
        Q \;, \hs{5} 
 \Psi = Q ({\bf 1}_M, \psi) \;, \hs{5}
 Q = \sqrt b ({\bf 1}_M + \psi\ph)^{-\1{2}} \;, 
 \label{fixing2}
\eeq
with $\ph$ and $\psi$ again being
$[(N-M)\times M]$- and $[M\times (N-M)]$-matrix chiral 
superfields, respectively. 
In this case, the superpotential is given by 
\beq
 W = b \sum_a m_a \tr \left[
    H_a \begin{pmatrix}
          {\bf 1}_M \cr \ph
        \end{pmatrix} 
    ({\bf 1}_M + \psi\ph)^{-1}
    ({\bf 1}_M, \psi)  
   \right] \;. \label{superpot2}
\eeq

\medskip
We can find the bundle structure 
of the manifold as follows.
First, let us consider the $b=0$ case. 
Then, setting $\psi=0$, the \kahler potential becomes 
\beq
 K|_{\psi=0} = c\, \tr \log (1+\ph\dagg\ph) \;, 
\eeq
which is that of the 
Grassmann manifold. 
Therefore, $\ph$ parameterizes the base Grassmann manifold, 
whereas $\psi$ parameterizes
 the cotangent space as the fiber\cite{LR}, 
with the total space being the cotangent bundle 
over the Grassmann manifold $T^* G_{N,M}$.
Next, consider the case $b\neq 0$. 
In the case of $T^*{\bf C}P^{N-1}$ with $M=1$, 
the base manifold is embedded by 
$\ph = \psi\dagg$~\cite{AF3}.\footnote{
The embedding $\ph = \psi\dagg$ should hold for 
a matrix with any value of $M$, 
although we have not yet proved this.
}

The tri-holomorphic isometry group of this manifold is 
$G = SU(N)$, whereas the isotropy group is 
$H = SU(M) \times SU(N-M) \times U(1)$, 
since the manifold is $T^* (G/H)$. 
In each case $b=0$ and $b\neq 0$, 
the action of $g \in SU(N)$ is found to be as follows:
\beq
 \Phi \to \Phi' = g \Phi h^{-1}(\ph,\psi) \;,\hs{5}
 \Psi \to \Psi' = h (\ph,\psi) \Psi g^{-1}\;. 
  \label{global-sym2}
\eeq
Here, $h (\ph,\psi) = e^{- i \Lambda(\ph,\psi)}$ 
is an induced gauge transformation called a 'compensator'. 
This is needed to put $g \Phi$ and $\Psi g^{-1}$ 
back into the gauge slice, 
because they are not in general in 
the gauge fixing condition (\ref{fixing1}) or (\ref{fixing2}). 
This makes the (base) manifold a coset space $G/H$. 
Under the global $SU(N)$ transformation, 
the \kahler potential undergoes the \kahler transformation
\beq
 K \to K' &=& K+
 c{\rm tr}\log (h^{-1}(\varphi,\psi))^\dagger 
 +c{\rm tr}\log h^{-1}(\varphi,\psi)\nonumber \\
 &=& K +  i c\, \tr \Lambda(\ph,\psi) 
 - i c\, \tr \Lambda\dagg(\ph\dagg,\psi\dagg) \;,
\eeq
whereas the \kahler metric is invariant. 
The tri-holomorphic Killing vectors for
the isometry $SU(N)$ can be calculated from 
infinitesimal transformations of (\ref{global-sym2}).

\medskip
Here we give several more comments.
\begin{enumerate}
\item
There exists a manifest duality between the 
two theories with the
$U(M)$ and $U(N-M)$ gauge symmetries 
and the same flavor $SU(N)$ symmetry. 
This results directly from the duality in 
the base Grassmann manifold $G_{N,M} \simeq G_{N,N-M}$. 

\item
For $M=1$ ($M=N-1$), namely
 for the $U(1)$ [$U(N-1)$] gauge symmetry,
 this model reduces to 
 $T^* {\bf C}P^{N-1} \simeq T^* G_{N,1} 
 ~(\simeq T^* G_{N,N-1})$, \cite{Ca} 
 which is discussed in detail in Ref.~\citen{ANNS}. 
Moreover, if $N=2$, the manifold $T^*{\bf C}P^1$ 
 is the Eguchi-Hanson space~\cite{EH}.

~~There are nontrivial models that do not reduce to 
 $T^*{\bf C}P^{N-1}$ target manifold. 
The nontrivial model with the lowest dimensions 
 of the target manifold is the case with $N=4, M=2$. 
In this case, the manifold is 
 $T^* G_{4,2} = T^* [SU(4)/ SU(2) \times SU(2) \times U(1)] 
 = T^* [SO(6)/SO(4) \times U(1)] \equiv T^* Q^4$,
 in which the base manifold $Q^4$ 
 is called the Klein quadric space.\footnote{
For diffeomorphisms of base manifolds, see Ref. \citen{HKN}.
} 

\end{enumerate}

\section{Vacuum structure}
In the massless case, $m_a=0$, 
 the moduli space of the vacua is 
 $T^*G_{N,M}$ itself 
 (the Higgs branch of ${\cal N}=2$ SUSY QCD~\cite{APS,AP}), 
 since the low energy effective theory has no potential term. 
However, once the mass term is switched on,
 most vacua are lifted up, leaving 
 some discrete points 
 as vacua.
First we consider the simpler case of $T^*{\bf C}P^{N-1}$ 
and then that of $T^*G_{N,M}$. 

\subsection{Vacua in the massive $T^* {\bf C}P^{N-1}$ model}
In this subsection, we investigate
$T^*{\bf C}P^{N-1} = T^* G_{N,1}$ with $M=1$. 
Without loss of generality we consider the case
$b=0$ with $c\neq 0$. 
In this case,
 the dynamical matrix fields are 
column and row vectors like 
$\ph^T = (\ph^1,\cdots,\ph^{N-1})$ and 
$\psi = (\psi^1,\cdots,\psi^{N-1})$.

The superpotential given in (\ref{superpot1}) becomes
\beq
 W = \sum_a m_a \tr \left[H_a 
       \begin{pmatrix}
           - \psi\cdot \ph & \psi \cr
          - \ph (\psi \cdot \ph) & \ph \otimes \psi
       \end{pmatrix} 
      \right] \;.
\eeq
We choose $H_a$ $(a=1,\cdots ,N-1)$ as 
\beq
 H_a =  \1{\sqrt{a(a+1)}} 
        \;{\rm diag.}\; (1,\cdots,1, -a, 0,\cdots,0)\;,
\eeq
where $-a$ is the ($a+1$)-th component, 
with the normalization given by 
the trace $\tr (H_a H_b) = \delta_{ab}$.
Then the superpotential can be calculated as
\beq
 W = - \sum_a M_a \psi^a \ph^a \;, \hs{5}
 M_a \equiv \sqrt{a \over a+1} m_a 
           + \sum_{b=1}^a {m_b \over \sqrt{b(b+1)}} \;. 
 \label{superpotCP}
\eeq
Therefore the derivatives of $W$ with respect to the fields are
\beq
  \del_{\ph^a} W = - M_a \psi^a \;, \hs{5}
 \del_{\psi^a} W = - M_a \ph^a \; \hs{5} (\mbox{no sum}) \;.
\eeq
These vanish only at the origin, $\ph= \psi^T =0$, 
 provided that $M_a\neq 0$.
This is the only vacuum in the regular region of 
 these coordinates, because the metric is regular there.

However, this model contains more vacua, 
 because the entire manifold is covered by several 
 coordinate patches, and a vacuum exists at the origin of each
 coordinate patch. 
To see this, we temporarily concentrate on the base 
${\bf C}P^{N-1}$.
We consider the fields before the gauge fixing, 
$\Phi \equiv \phi^A = (\phi^1, \cdots, \phi^N)^T$ 
($A= 1,\cdots,N$), 
called the homogeneous coordinates,  
in which we need an identification 
 by the gauge transformation 
 $\phi^A \sim e^{i\Lam} \phi^A$ to realize the space 
 ${\bf C}P^{N-1}$. 
In the region satisfying $\phi^1 \neq 0$, we can choose 
a patch $\ph^i = \phi^{i+1}/\phi^1$ $(i= 1,\cdots,N-1)$,
which was used in Eq.~(\ref{fixing1}). 
Here, let us write these coordinates as 
$\ph_{(1)}^i = \phi^{i+1}/\phi^1$. 
In the same way, in the region satisfying
$\phi^A \neq 0$, 
we can choose the $A$-th patch defined by
\beq  
 \ph_{(A)}^i 
  = \begin{cases}
           \phi^i/\phi^A,    \hs{5} (1 \leq i \leq A-1)  \cr
           \phi^{i+1}/\phi^A. \hs{5} (A \leq i \leq N-1)
    \end{cases}
  \;.
\eeq
We thus have $N$ sets of patches $\{\ph_{(A)}^i \}$, 
which is sufficient
 to cover the whole base manifold. 
Corresponding to each patch for the base space, 
we manifestly have an associated patch for 
the fiber tangent space $\{\psi_{(A)}^i \}$, from 
Eq.~(\ref{fixing1}).  
These sets of coordinates, 
$\{\ph_{(A)}^i, \psi_{(A)}^i\}$, 
are sufficient to cover the whole $T^*{\bf C}P^{N-1}$.
For each patch, 
the origin $\ph_{(A)}^i = \psi_{(A)}^i = 0$ 
is a vacuum. 
Therefore, the number of discrete vacua for 
the massive $T^*{\bf C}P^{N-1}$ model is $N$.
This result was  
 first obtained in Ref. \citen{GTT2}.

To investigate solitons like BPS walls, 
their junction and lumps, 
it may be better to consider the problem 
in one coordinate patch. 
The other vacua appear in one patch 
as the coordinate singularities of 
the metric 
in infinities of the coordinates 
rather than the stationary points of the 
superpotential~\cite{NNS}. 
To see this, we again consider only the base ${\bf C}P^{N-1}$.
We study how the $A$-th vacuum ($A\neq 1$) 
in the origin of 
the $A$-th coordinate patch is mapped into 
the first patch.
The $A$-th vacuum is represented by 
$\ph_{(A)}^i = 0$ 
or $\phi^B/\phi^A = 0$ for all $B~(\neq A)$.
In the first coordinate patch, this point is mapped 
 to an infinite point represented by 
\beq
 \ph_{(1)}^{A-1} \to \infty \;,\hs{5}  
 \ph_{(1)}^i / \ph_{(1)}^{A-1} \to 0 
   \;\; (i \neq A-1), 
\eeq
which appears to be a runaway vacuum in this patch. 
Hence, the origin and $N-1$ infinities are vacua
in each coordinate patch~\cite{NNS}. 
To summarize, 
if we include runaway vacua, 
one patch is sufficient to describe 
soliton solutions.\footnote{The domain wall solution connects 
the origin and the infinity in this parameterization 
for the $T^*{\bf C}P^1$ case \cite{ANNS}.
} 
However, note that the terminology ``runaway`` is 
merely a coordinate-dependent concept, 
because a runaway vacuum in one coordinate patch
is a true vacuum in another coordinate patch.

\medskip
We can also study the vacua without referring to 
the local coordinate patches. 
We concentrate on the base ${\bf C}P^{N-1}$ once again. 
A point in ${\bf C}P^{N-1}$ 
 corresponds to a complex line through the origin 
 in ${\bf C}^N$ with homogeneous coordinates $\phi^A$, 
 because the gauge transformation corresponds to 
 the equivalence relation  
 $\phi^A\sim e^{i\Lam}\phi^A$.  
The first vacuum is expressed in the 
 region satisfying $\phi^1 \neq 0$ 
 by the relations 
 $\ph_{(1)}^i = \phi^{i+1}/\phi^1 =0$ $(i=1,\cdots,N-1)$, 
 namely $\phi^{i+1} = 0$. 
Therefore, the first vacuum corresponds to 
 the $\phi^1$-axis. 
Similarly, the $A$-th vacuum corresponds to 
 the $\phi^A$-axis. 
In this way, each vacuum is simply expressed by  
 one orthogonal axis in ${\bf C}^N$. 
Note that each axis is invariant under 
 the $U(1)^{N-1}$ transformation of $H_a$, and hence
 it is a fixed point of this transformation. 

If we take $N$ orthogonal normalized basis vectors $e_A$ 
 [with $(e_A)^*\cdot e_B = \delta_{AB}$], 
 whose components are given by 
\beq
 (e_A)^B =\delta_A^B, \label{basis}
\eeq
then for any 
 complex line in ${\bf C}^N$, there exists a unit
 vector $e' = \sum_{A=1}^N a^A e_A = U e_1$ 
 that spans this line,
 where each $a^A$ is a complex number with 
 $\sum_A |a^A|^2 = 1$ and 
 $U$ is a unitary matrix, $U \in U(N)$. 
Each of the $N$-vacua found above corresponds to 
 a different one of the vectors $e_A$ ($A=1,\cdots,N$) 
 (with a zero value of the cotangent space $\psi=0$).

\begin{figure}
\begin{center}
\leavevmode
  \epsfysize=9.0cm
  \epsfbox{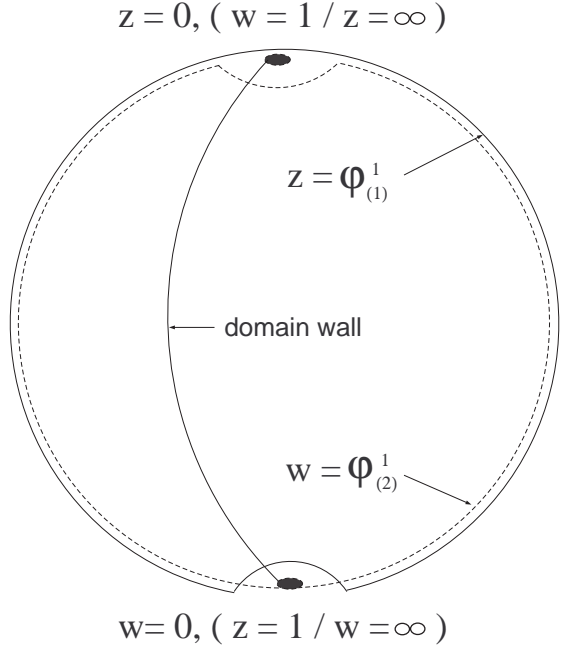} \\ 
\caption{
The base manifold of $T^*{\bf C}P^1$ and vacua.}
\label{fig1}
\end{center}
\begin{small}
Corresponding to two gauge fixing conditions, 
we have two coordinates, $z$ and $w$, 
covering $S^2$, except for the south (S) and north (N) poles, 
respectively.
The origins of $z$ and $w$ ( N and S, respectively) 
are both vacua. 
The domain wall solution, 
approaching these two vacua at spatial infinities,
is mapped to a trajectory connecting N and S in $S^2$. 
\end{small}
\end{figure}

\medskip
\underline{Example:  
the Eguchi-Hanson space}~\cite{EH}.
The simplest model is the Eguchi-Hanson space, 
$T^*{\bf C}P^1$ ($N=2,M=1$). 
This model has two discrete vacua and admits 
 a typical domain wall solution~\cite{AT,ANNS}.
The vacua are located at the north and south poles 
 of the base ${\bf C}P^1 \simeq S^2$ (see Fig.~\ref{fig1}).
Corresponding to the two gauge fixing conditions 
 $\Phi = \begin{pmatrix} 1 \cr z \end{pmatrix}$ 
 and $\Phi = \begin{pmatrix}w \cr 1 \end{pmatrix}$, 
 we have the two coordinate patches 
 $z \equiv \ph^1_{(1)} = \phi^2/\phi^1$ and
 $w \equiv \ph^1_{(2)} = \phi^1/\phi^2$, 
 where we have $z = 1/w$. 
Two vacua are given by $z = 0$ and $w=0$. 
 The second (first) vacuum, $w=0$ ($z=0$), is mapped to 
 $z= \infty$ ($w=\infty$) in the first (second) patch, 
 which looks like a runaway vacuum.
In homogeneous coordinates, 
these correspond to 
$\left< \Phi \right> = \begin{pmatrix} 1 \cr 0 \end{pmatrix} \equiv e_1$ 
and 
$\left< \Phi \right> = \begin{pmatrix} 0 \cr 1 \end{pmatrix} \equiv e_2$, 
respectively, with $\left< \Psi \right> = (0,0)$.
Also, in a coordinate independent manner, 
these two vacua correspond to the 
$\phi^1$ and $\phi^2$ axes, spanned by 
$e_1$ and $e_2$, respectively.

\medskip
Before closing this subsection, 
we consider the case $b \neq 0$.
In this case,  
 The superpotential (\ref{superpot2}) 
 can be calculated, and we have
\beq
&& W = {b \over 1+\psi\cdot\ph} 
    \left( L + \sum_{a=1}^{N-1} N_a \psi^a \ph^a \right) \,, \non
&& L \equiv \sum_{a=1}^{N-1} {m_a \over \sqrt{a(a+1)}} \;,\hs{5}
   N_a \equiv - \sqrt{a \over a+1} m_a 
  + \sum_{b=a+1}^{N-1} {m_b \over \sqrt{b(b+1)}} = L- M_a\;,
\eeq
with $M_a$ defined in (\ref{superpotCP}).
The derivatives of $W$ are given by
\beq
 \del_{\ph^a} W 
  &=& - {b \psi^a \over (1+ \psi\cdot\ph)^2} 
  \left[L - N_a + \sum_{b=1}^{N-1} (N_b - N_a) \psi^b \ph^b \right] \non
  &=& - {b \psi^a \over (1+ \psi\cdot\ph)^2} 
  \left[M_a - 
  \sum_{b=1}^{N-1} (M_b - M_a) \psi^b \ph^b \right]
    \; , \non
 \del_{\psi^a} W 
   &=& (\psi^a \leftrightarrow \ph^a) \;,
\eeq
where the arrow in the second equation indicates 
the exchange of quantities in the first equation.
The origin, $\ph^a=\psi^a =0$, in each patch is a vacuum. 
There is no vacuum other than these $N$ vacua.
The number of vacua should coincide with that in
 the case with $b=0$ and $c\neq 0$, 
because they are connected by the R-symmetry, and 
the physics does not depend on the difference.

\subsection{Vacua in the massive $T^* G_{N,M}$ model}
To determine the vacua of the $T^* G_{N,M}$ model,
we consider the case $b=0$ and $c\ \neq 0$, again 
without loss of generality. 
We write the matrix
 $\ph$ as $(\ph_{i \alpha})$ and 
 $\psi$ as $(\psi_{\alpha i})$, 
 with the indices $i=1,\cdots, N-M$ and $\alpha = 1,\cdots,M$. 
The superpotential given in Eq.~(\ref{superpot1}) 
 can be calculated as
\beq
 && W = - \sum_{\alpha =1}^M \sum_{i=1}^{N-M} 
     M_{\alpha i} \ph_{i \alpha} \psi_{\alpha i} \;, \non
 && M_{\alpha i} \equiv 
   \sqrt{i+M-1 \over i+M} m_{i+M-1} 
 - \sqrt{\alpha-1\over \alpha} m_{\alpha - 1} 
 + \sum_{a = \alpha}^{i+M-1} {m_a \over \sqrt {a(a+1)}} \;,                
 \label{superpotGr}
\eeq
where we have set $m_0 \equiv 0$. 
For the case $M=1$ ($\alpha=1$), 
 this reduces to Eq.~(\ref{superpotCP}) for 
 $T^* {\bf C}P^{N-1}$. 
The derivatives of superpotential given in (\ref{superpotGr}) 
 with respect to the fields are
\beq
 \del_{\ph_{i \alpha} } W 
   = - M_{\alpha i} \psi_{\alpha i} \;,\hs{5} 
 \del_{\psi_{\alpha i} } W 
   = - M_{\alpha i} \ph_{i \alpha} \; \hs{5} (\mbox{no sum}) \;.
\eeq
Therefore, the origin of these coordinates, $\ph = \psi^T = 0$,
 is a vacuum, provided that $M_{ai}\neq 0$.
This is the unique vacuum in 
 the finite region of these coordinates 
 where the metric is regular.
The number of vacua in this model is equal to
 the number of coordinate patches, 
 as in the $T^*{\bf C}P^{N-1}$ case.
In the first coordinate patch, 
 we have chosen the first $M$ row 
 vectors in $\Phi$ as the unit matrix, as in 
 Eqs.~(\ref{fixing1}) or (\ref{fixing2}). 
The other coordinate patches are obtained by 
 making other choices of the gauge fixing conditions, 
 with which the other sets of $M$ row vectors 
 in $\Phi$ form the unit matrix.
The number of such coordinate systems is 
$_N C_M = N!/[M!(N-M)!]$. 
They are independent and  
 sufficient to cover the entire manifold.
Therefore, 
 this model has $N!/[M! (N-M)!]$ vacua.
This number is invariant 
under the duality between 
$U(M)$ and $U(N-M)$ gauge groups.
It also reduces correctly to $N$ for $T^*{\bf C}P^{N-1}$ 
when $M=1$ or $M=N-1$.

As in the $T^* {\bf C}P^{N-1}$ case, 
we can understand the vacua of $T^* G_{N,M}$ 
without local coordinates.
A point in the base $G_{N,M}$ corresponds to 
 an $M$-dimensional complex hyperplane ($M$-plane) 
 through the origin in ${\bf C}^N$.
The vacua found above correspond to  
 mutually orthogonal $M$-planes spanned by
 $M$ arbitrary sets of axes chosen from the $N$ axes. 
Therefore, the total number of vacua 
 is $_N C_M = N!/[M!(N-M)!]$. 
Because the $M$-planes of vacua are invariant under 
 $U(1)^{N-1}$ generated by $H_a$, 
 the vacua are fixed points.

Choosing the basis (\ref{basis}) in ${\bf C}^N$, 
 a point in $G_{N,M}$ 
 expressed as an $M$-plane in ${\bf C}^N$ can be 
 spanned by a set of $M$ unit vectors, 
\beq
  (e_i)' = U e_i \;, \hs{5} 
\eeq
where $i = 1,\cdots,N-M$ and 
 $U$ is a unitary matrix, $U \in U(N)$. 
The vacua of mutually orthogonal $M$-planes are 
 spanned by $M$ arbitrary sets of 
 basis vectors among the orthogonal $N$-basis. 

The duality becomes manifest in this framework. 
We can represent a point in $G_{N,M}$ by 
 an $(N-M)$-plane that is the complement of an $M$-plane.

\medskip
\underline{Example: the cotangent bundle over the Klein quadric}.
We now consider the  
 example of the Klein quadric $T^* G_{4,2} = T^* Q^4$ 
($N=4$ and $M=2$). 
In this case, 
 there exist six coordinate systems $\ph^{(A)}_{i\alpha}$ 
 ($A=1,\cdots,6$) for the base manifold 
 corresponding to six choices of gauge fixing, 
 given by
\beq
 \Phi &=& 
 \begin{pmatrix} 1 & 0 \cr 0 & 1 \cr 
         \ph^{(1)}_{11} & \ph^{(1)}_{12} \cr 
         \ph^{(1)}_{21} & \ph^{(1)}_{22}\end{pmatrix} \,, \hs{3} 
 \begin{pmatrix} 1 & 0 \cr \ph^{(2)}_{11} & \ph^{(2)}_{12} \cr 
           0 & 1 \cr \ph^{(2)}_{21} & \ph^{(2)}_{22}\end{pmatrix} \,, \hs{3} 
 \begin{pmatrix} 1 & 0 \cr \ph^{(3)}_{11} & \ph^{(3)}_{12} \cr 
           \ph^{(3)}_{21} & \ph^{(3)}_{22} \cr 0 & 1\end{pmatrix} \,, \hs{3} \non
&& \begin{pmatrix} \ph^{(4)}_{11} & \ph^{(4)}_{12} \cr 1 & 0 \cr 
             0 & 1 \cr \ph^{(4)}_{21} & \ph^{(4)}_{22}\end{pmatrix} \,, \hs{3} 
 \begin{pmatrix} \ph^{(5)}_{11} & \ph^{(5)}_{12} \cr 1 & 0 \cr 
           \ph^{(5)}_{21} & \ph^{(5)}_{22} \cr 0 & 1\end{pmatrix} \,, \hs{3} 
 \begin{pmatrix} \ph^{(6)}_{11} & \ph^{(6)}_{12} \cr 
           \ph^{(6)}_{21} & \ph^{(6)}_{22} \cr 1 & 0 \cr 0 & 1\end{pmatrix} \, . 
  \label{Klein}
\eeq
Together with the corresponding coordinates $\psi^{(A)}_{\alpha i}$ 
 for the cotangent space in Eq.~(\ref{fixing1}), 
 these six sets of coordinate systems are 
 sufficient to cover the entire manifold.
Therefore this model has the six vacua given by
\begin{eqnarray}
 &&\left<\Phi\right> = 
 \begin{pmatrix} 1 & 0 \cr 0 & 1 \cr 0 & 0 \cr 0 & 0 \end{pmatrix} \,, \hs{3} 
 \begin{pmatrix} 1 & 0 \cr 0 & 0 \cr 0 & 1 \cr 0 & 0\end{pmatrix} \,, \hs{3} 
 \begin{pmatrix} 1 & 0 \cr 0 & 0 \cr 0 & 0 \cr 0 & 1\end{pmatrix} \,, \hs{3} 
 \begin{pmatrix} 0 & 0 \cr 1 & 0 \cr 0 & 1 \cr 0 & 0\end{pmatrix} \,, \hs{3} 
 \begin{pmatrix} 0 & 0 \cr 1 & 0 \cr 0 & 0 \cr 0 & 1\end{pmatrix} \,, \hs{3} 
 \begin{pmatrix} 0 & 0 \cr 0 & 0 \cr 1 & 0 \cr 0 & 1\end{pmatrix} \,, 
 \nonumber \\ 
 && \label{vac-Klein}
\end{eqnarray}
which are the respective for the six choices
 given in (\ref{Klein}),
 with $\left<\Psi\right>=0$.
The set of two column vectors in each matrix 
 in Eq.~(\ref{vac-Klein}) constitutes 
 a set of orthogonal basis vectors
 $e_i$ chosen from the four basis vectors.

\medskip
In the case $b \neq 0$, 
the superpotential (\ref{superpot2}) is 
\beq
 W &=& b \sum_{a=1}^{N-1} \sum_{n=0}^{\infty} (-1)^n  
    m_a \tr \left[ H_a 
     \begin{pmatrix} (\psi\ph)^n     & (\psi\ph)^n \psi \cr
               \ph (\psi\ph)^n & (\ph \psi)^{n+1} \end{pmatrix}
      \right] \non
 &=& b \sum_{a=1}^{N-1} \sum_{n=0}^{\infty} (-1)^n  
    m_a \tr \left[ H_a 
     \begin{pmatrix} (\psi\ph)^n & 0 \cr
                         0 & (\ph \psi)^{n+1} \end{pmatrix}
      \right] \;, \label{superpot3}
\eeq
where the last equality holds because the matrices
 $H_a$ are diagonal.
 Similarly to the $T^*{\bf C}P^{N-1}$ case, 
 the origin $\ph = \psi^T =0$ of each patch is a vacuum 
 and we cannot have any other vacua.

\section{Massive HK quotient in the Wess-Zumino gauge}
In the previous sections, we eliminated 
 ${\cal N}=2$ vector multiplets 
 (as auxiliary superfields)
 without taking the Wess-Zumino gauge. 
In this section, we take the Wess-Zumino gauge, 
 derive the bosonic action, and investigate vacua.

\subsection{Lagrangian in the Wess-Zumino gauge}
It is difficult to calculate the scalar potential 
 without taking the Wess-Zumino gauge, 
 because the inverse metric is difficult to obtain.
On the other hand, in the Wess-Zumino gauge,
 the scalar potential can be directly obtained, 
 while the \kahler potential and the superpotential 
 cannot be calculated easily.
We calculate the scalar potential 
 in this subsection. 
We represent the lowest components in the superfields using the
 same letters as we have used for the corresponding
 superfields. 

In the Wess-Zumino gauge, 
the bosonic action of (\ref{linear}) is given by 
\begin{equation}
 {\cal L}_{\rm boson} = 
 {\cal L}_{\rm kin} + 
 {\cal L}_{\rm constr} + 
 {\cal L}_{\rm pot},  
\label{eq:N1U1boson}
\end{equation}
in which each term is given by 
\beq
 {\cal L}_{\rm kin} 
&=& - \tr \left[ 
      (\partial_{\mu} \Phi\dagg - i v_{\mu} \Phi\dagg  ) 
      (\partial^{\mu} \Phi+ i \Phi v^{\mu} ) \right] 
 - \tr \left[ 
      (\partial_{\mu} \Psi - i v_{\mu} \Psi )
      (\partial^{\mu} \Psi\dagg + i \Psi\dagg v^{\mu} ) 
      \right] \non
&=& 
- \tr (\del_{\mu}\Phi\dagg \del^{\mu}\Phi 
     + \del_{\mu}\Psi \del^{\mu}\Psi\dagg ) 
+ i \tr \left[ v^{\mu} (
\Phi^{\dagger} 
\dellr_{\mu} \Phi 
 + \Psi \dellr_{\mu}\Psi\dagg) \right] \non
&&
-  \tr \left[ v^{\mu} v_{\mu}
   (\Phi^{\dagger} \Phi + \Psi \Psi\dagg) \right], \\
 {\cal L}_{\rm constr} & = & 
\tr [ D (\Phi^{\dagger} \Phi 
        - \Psi \Psi\dagg - c {\bf 1}_M ) ] 
+
 \tr [ F_\Sigma (\Psi \Phi - b {\bf 1}_M ) + {\rm c.c.} ], 
 \label{eq:N1U1constr-lag} \\
 {\cal L}_{\rm pot} & = & 
\tr ( F_\Phi^{\dagger} F_\Phi + F_\Psi F_\Psi\dagg) \non
&& 
+ \left[ \tr \{ \Sigma(F_{\Psi} \Phi + \Psi F_{\Phi})\}
+ \sum_a m_a 
  \tr (F_{\Psi} H_a \Phi + \Psi H_a F_{\Phi}) 
 + {\rm c.c.} \right]\non
&
\equiv
& 
 -V(\Phi,\Psi,\Sigma),
\label{eq:N1U1pot}
\eeq
where we have defined $A \dellr_{\mu} B
\equiv A (\partial_{\mu} B) - (\partial_{\mu} A ) B$.

The equation of motion for the gauge field $v_{\mu}$ 
without the kinetic term in (\ref{eq:N1U1boson}) reads 
\beq
 \del {\cal L}/\del v_{\mu} 
 = - \{v_{\mu}, (\Phi\dagg\Phi + \Psi\Psi\dagg) \}
   + i (\Phi\dagg \dellr_{\mu} \Phi 
      + \Psi \dellr_{\mu}\Psi\dagg )
 = 0 \;.
 \label{eq:gauge-field}
\eeq
Here $\{~,~\}$ is an anti-commutator.
This equation can be solved by expanding fields 
 that takes values in the Lie algebra 
 in terms of the $M\times M$ matrix $T_A$ 
 of the fundamental representation of $U(M)$ generators 
\begin{eqnarray}
v_{\mu} & \equiv & v_{\mu}^A T_A, \qquad 
{\rm tr}\left(T_A T_B\right)= \delta_{AB}
\nonumber \\
\Phi^\dagger \Phi + \Psi \Psi^\dagger & \equiv & 
A^A T_A, \qquad 
\tilde A_{AB}\equiv {1 \over 2} 
{\rm tr} \left(\left\{ T_A, 
T_B\right\} T_C \right) A^C, 
\nonumber \\
 i (\Phi\dagg \dellr_{\mu} \Phi 
      + \Psi \dellr_{\mu}\Psi\dagg )
 & \equiv & B_{\mu}^A T_A .
\end{eqnarray}
Then we can express the gauge fields $v_{\mu}$ 
 in terms of the dynamical scalar fields, 
 solving the equation of motion (\ref{eq:gauge-field})
 as 
\begin{equation}
v_{\mu}^A = \frac{1}{2} \left(\tilde A^{-1}\right)_{AB} B_{\mu}^B .  
  \label{U(1)quotient}
\end{equation}
If we eliminate the gauge field using this algebraic 
 equation of motion, we obtain the kinetic term for 
 hypermultiplets as 
\begin{eqnarray}
 {\cal L}_{\rm kin}  =  
- \tr (\del_{\mu}\Phi\dagg \del^{\mu}\Phi 
     + \del_{\mu}\Psi \del^{\mu}\Psi\dagg ) 
+ {1 \over 4}  B^{A\mu} \left(\tilde A^{-1}\right)_{AB} 
B_{\mu}^B 
. 
\label{eq:N1U1kin}
\end{eqnarray}

If we integrate the Lagrange multiplier fields 
 $D$ and $F_\Sigma$ in (\ref{eq:N1U1constr-lag}), 
 we obtain three real constraints 
\begin{equation}
\Phi\dagg \Phi - \Psi\Psi\dagg = c {\bf 1}_M,  \qquad 
\Psi\Phi = b {\bf 1}_M  .
\label{eq:N1U1constr}
\end{equation}
The left-hand sides of these relations constitute 
 the triplet of the moment map (Killing potential) 
 for the $U(M)$ gauge symmetry. 
These values are fixed to
 the FI parameters, 
 and the hyper-\kahler quotient is obtained 
 together with the $U(M)$ quotient given in (\ref{U(1)quotient}). 

The Lagrangian (\ref{eq:N1U1pot}) gives the following 
 algebraic equations of motion for 
 the auxiliary fields $F_\Phi$ and $F_\Psi$: 
\begin{eqnarray}
 F_\Phi^{\dagger} 
  = - \Sigma \Psi - \sum_a m_a \Psi H_a , \qquad 
 F_\Psi^{\dagger} 
  = - \Phi \Sigma - \sum_a m_a H_a \Phi .  
  \label{eq:EOMF}
\end{eqnarray}
After eliminating the auxiliary fields $F_\Phi$ and $F_\Psi$ 
 using these algebraic equations of motion,  
 we obtain the potential term 
$V$ in Eq.(\ref{eq:N1U1pot}) as 
\begin{eqnarray}
V(\Phi,\Psi,\Sigma) &=& 
\tr ( F_\Phi^{\dagger} F_\Phi + F_\Psi F_\Psi\dagg) 
 \nonumber \\
     &=& 
\tr \left( (\Sigma \Psi + \sum_a m_a \Psi H_a)
(\Psi^\dagger\Sigma^\dagger + \sum_b m_b^* H_b \Psi^\dagger)
\right. \nonumber \\
     & & 
+ \left.
 (\Sigma^\dagger \Phi^\dagger 
 + \sum_a m_a^* \Phi^\dagger H_a)
(\Phi\Sigma + \sum_b m_b H_b \Phi)\right) 
 \nonumber \\
     &=& 
\tr \left( \Sigma^\dagger \Sigma \Psi \Psi^\dagger 
+ \sum_{ab} m_a m_b^* \Psi H_a H_b \Psi^\dagger\right)
 \nonumber \\
     & & 
+ \, \tr \left(\Sigma\Sigma^\dagger \Phi^\dagger \Phi 
 + \sum_{a b} m_a^* m_b \Phi^\dagger H_a H_b \Phi\right) 
 \nonumber \\  
    && + \sum_a 
\tr \left(\left(m_a^* \Sigma + m_a \Sigma^\dagger\right) 
\left(\Phi^\dagger H_a \Phi + \Psi H_a \Psi^\dagger\right)
\right) .
\label{eq:N1U1pot2}
\end{eqnarray}
Next, to eliminate $\Sigma$, we define the following quantities: 
\begin{eqnarray}
\Sigma &\equiv & \Sigma^A T_A, \qquad 
\Sigma^\dagger = \Sigma^{A*} T_A, 
 \nonumber \\
\Psi \Psi^\dagger  & \equiv & C^A T_A, \qquad 
\tilde C_{AB}\equiv \tr \left(T_A T_B T_C\right) 
C^C, 
  \nonumber \\
\Phi^\dagger \Phi &\equiv & D^A T_A, \qquad 
\tilde D_{AB}\equiv \tr \left(T_B T_A T_C\right)
D^C,
 \nonumber \\
&&\sum_a m_a^* \left(\Phi^\dagger H_a \Phi+ 
\Psi H_a \Psi^\dagger\right)\equiv E^A T_A.
\end{eqnarray}
Because the equation of motion for $\Sigma$ is purely 
 algebraic, we can eliminate it as 
\begin{equation}
\Sigma^A = 
-\left(\left(\tilde C+\tilde D\right)^{-1}\right)_{AB}
E^{B*}, 
\qquad 
\Sigma^{A*} = - E^{B}
\left(\left(\tilde C+\tilde D\right)^{-1}\right)_{BA}.  
\label{eq:sigma}
\end{equation}
After eliminating the auxiliary fields $\Sigma$ 
 in the ${\cal N}=2$ vector multiplet $(V, \Sigma)$ 
 using its algebraic equation of motion, 
 we obtain the potential in terms of dynamical scalar 
 fields, 
\begin{eqnarray}
V(\Phi, \Psi) &= &
\tr \left( 
 \sum_{ab} \left(m_a m_b^* \Psi H_a H_b \Psi^\dagger
 + m_a^* m_b \Phi^\dagger H_a H_b \Phi\right)\right) 
 \nonumber \\  
&& 
- E^{A}
\left(\left(\tilde C+\tilde D\right)^{-1}\right)_{AB} E^{B*}.  
\label{eq:potential}
\end{eqnarray}

\subsection{Vacua in the massive $T^* {\bf C}P^{N-1}$ model}
We first focus on the vacua of the $T^* {\bf C}P^{N-1}$ 
 model (the $M$=1 case).
Here, we are interested in the SUSY vacua.
SUSY vacua correspond to vanishing auxiliary fields 
 $F_{\Phi}$ and $F_\Psi$, i.e., 
\begin{eqnarray}
&& 0 = - F\dagg_{\Phi^i} 
     = \Psi_i(\Sigma + \tilde{m}_i ), \label{WZvac1}\\
&& 0 = - F\dagg_{\Psi_i} 
     = ( \Sigma  + \tilde{m}_i ) \Phi^i, \label{WZvac2}
\end{eqnarray}
and the constraints
\begin{eqnarray}
&& 0
   = \Phi_i\dagg \Phi^i - \Psi_i \Psi^{\dagger i}- c,  \label{WZvac3}\\
&& 0
   = \Psi_i\Phi^i - b \,,\label{WZvac4} 
\end{eqnarray}
where we have defined $\tilde{m}_i$ by 
 $\sum_a m_a H_a={\rm diag.}(\tilde{m}_1,\tilde{m}_2,
 \dots,\tilde{m}_N)$, 
 with $\sum_{i=1}^N \tilde{m}_i = 0$.
We assume a generic mass $\tilde{m}_i\neq \tilde{m_j}$
\footnote{
 This assumption is the same as the assumption 
 $M_a\neq 0$, which was made 
 in the study of the vacua in $\S$3.1. 
Indeed, the condition $\tilde{m}_i\neq\tilde{m}_j$ for any 
 combination of $i$ and $j$ ($i \neq j$) is equivalent to 
 the condition $M_a\neq 0$ for 
 all $a$ in all patches.  
If $\tilde{m}_i=\tilde{m}_j$ holds for a pair $i$ and $j$, 
 there is a patch where $M_a=0$ for one $a$.
The vacuum is then not localized at the origin of the patch, 
 and hence continuous vacua appear.
Therefore,
 the discussion in $\S$3.1 no longer holds.
For instance, in the $N=3$ case, the relation
 $M_a=\tilde{m}_3-\tilde{m}_a$ holds
 in the patch, and therefore we have $\Phi^t=(\varphi,1)^t$ and 
 $\Psi=(\psi,-\psi\varphi)$ instead of (\ref{fixing1}).
It is easily seen that $M_2=0$ if $\tilde{m}_2=\tilde{m}_3$, 
 while $M_1\neq 0$.} 
 and that either $b$ or $c$ has a nonzero value.
Equation (\ref{WZvac1}) implies $\Psi_i = 0$ or 
\beq
 \Sigma = - \tilde{m}_i, \label{WZvac}
\eeq
with $\Psi_i \neq 0$ for some $i~(= 1,2,\cdots,N)$. 
In the case $\Sigma \not= -\tilde m_i$ for all $i$, 
 we have $\Phi=0$ and $\Psi=0$, from Eqs.~(\ref{WZvac1}) 
 and (\ref{WZvac2}). 
However, $\Phi = \Psi^T = 0$ is inconsistent 
 with Eq.~(\ref{WZvac3}) for $c\neq 0$ or 
 with Eq.~(\ref{WZvac4}) for $b \neq 0$. 
Therefore Eq.~(\ref{WZvac}) must hold for some $i$, 
 and $\Phi^i$ and/or $\Psi_i$ can be nonzero for such $i$. 
Hence, there exist $N$ vacua labelled by $i=1,\cdots,N$. 
To determine the vacuum expectation values of 
 $\Phi^i$ and $\Psi_i$ for the $i$-vacuum, 
 let us assume $b=0$ and $c \neq 0$, without loss of generality.
Then, from Eq.~(\ref{WZvac4}) with $b=0$, 
 $\Phi^i$ or $\Psi_i$ must be zero.
Using Eq.~(\ref{WZvac3}), we obtain 
 $|\Phi^i| = \sqrt c$ and $\Psi_i =0$ if $c > 0$ 
 or $\Phi^i = 0$, and $|\Psi_i| = \sqrt{ -c}$ if $c < 0$. 
In the case $c=0$ and $b \neq 0$, 
we obtain $\Phi^i =\sqrt be^{i\theta}$ and 
 $\Psi_i = \sqrt be^{-i\theta}$ with 
 an arbitrary phase $\theta$. 
We thus have found $N$ discrete vacua.

\subsection{Vacua in the massive $T^* G_{N,M}$ model}
\label{vacuasection}
The analysis of vacua for the massive 
 $T^*{\bf C}P^{N-1}$ model in the previous subsection can be 
 generalized to the case of the massive $T^*G_{N,M}$ model.
The vacuum conditions in this case are given by 
\begin{eqnarray}
&& 0_{\alpha i} = - F_{\Phi \alpha i}^{\dagger}
     = (\Sigma_{\alpha}{}^\beta
     + \delta_{\alpha}{}^\beta \tilde{m}_i)\Psi_{\beta i}, 
\label{WZvac5}\\
&& 0^{i \alpha} = - F_{\Psi}^{\dagger i\alpha} 
     = \Phi^{i\beta} ( \Sigma_{\beta}{}^\alpha 
       + \delta_{\beta}{}^\alpha \tilde{m}_i), 
\label{WZvac6}
\end{eqnarray}
along with the constraints
\begin{eqnarray}
&& 0_\alpha^{~\beta} = \Phi\dagg_{\alpha i} \Phi^{i\beta} 
 - \Psi_{\alpha i}\Psi^{\dagger i\beta} 
 - c \delta_\alpha^{~\beta}, \label{WZvac7}\\
&& 0_\alpha^{~\beta} = \Psi_{\alpha i} \Phi^{i\beta} 
   - b \delta_{\alpha}{}^{\beta}\,,
   \label{WZvac8} 
\end{eqnarray}
where $\alpha,\beta=1,\cdots,M$ are gauge indices. 
Using the $U(M)$ gauge rotation, $\Sigma$ can be diagonalized:
\begin{eqnarray}
&\Sigma={\rm diag.}(\Sigma_1,\Sigma_2,\cdots,\Sigma_M),& \label{diag}\\
&\Sigma_1+\Sigma_2+\cdots+\Sigma_M=M \Sigma^0.&
\end{eqnarray}
To infer the consequences of Eqs.~(\ref{WZvac5}) and 
 (\ref{WZvac6}), we should consider two cases separately, 
 that in which 
 $\Sigma_\alpha+\tilde{m}_i\neq 0$ for some 
 $\alpha$ and all $i$ and that in which 
 $\Sigma_\alpha+\tilde{m}_i = 0$ for
 all $\alpha$ and some $i$.
In the former case, 
 we get $\Phi=0$ and $\Psi=0$. 
However, these are inconsistent with (\ref{WZvac7}) 
 for $c\neq 0$ and with (\ref{WZvac8}) for $b \neq 0$,
 and hence they do not represent a solution.
In the latter case, let us assume the generic case for
 masses, i.e. $\tilde m_i\not=\tilde m_j$.
\footnote{As in the case of the $T^*{\bf C}P^{N-1}$ model,
 it can be confirmed that 
 this assumption is the same as $M_{ai}\neq 0$ 
 used in $\S$3.2.}
The eigenvalue equation 
 (\ref{WZvac6}) for $\Sigma_\alpha$ gives 
 the eigenvalue $\Sigma_\alpha=-\tilde m_i$ 
 corresponding to the eigenvector formed 
 by the $\alpha$-th column vector, 
 $(\Phi^{1\alpha}, \cdots,  \Phi^{N\alpha})^T$, 
 with only a single nonvanishing element. 
Similarly, the eigenvalue equation 
 (\ref{WZvac5}) gives the same eigenvalue 
 $\Sigma_\alpha =-\tilde m_i$ corresponding to
 the eigenvector formed by the $\alpha$-th row vector, 
 $(\Psi_{\alpha 1}, \cdots, \Psi_{\alpha N})$, 
 with only a single nonvanishing element.
Explicitly, we have 
\begin{eqnarray}
\Phi=\left(
                \begin{array}{cccccc}
		  0           \\    
                  \vdots      \\
                  \Phi^{i\alpha} \\
                  0           \\
                  \vdots      
		\end{array}
               \right),~~~~~
\Psi=\left(
                \begin{array}{ccccc}
		 0      & \cdots & \Psi_{\alpha i} & 0 & \cdots 
		\end{array}
               \right)
. \label{eigen}
\end{eqnarray}
Since there are $N$ different masses $\tilde m_i$, 
 there are $N$ corresponding nontrivial eigenvectors 
 of the form given in (\ref{eigen}).
The diagonal $\Sigma_\alpha$ should be equal to one such 
 mass, $-\tilde m_i$. 
We have $M$ different diagonal elements 
 $\Sigma_\alpha$ ($\alpha=1, \cdots, M$). 
Therefore, we obtain ${}_NC_M$ possibilities of different 
 vacua, counting a vacua related by relabelling of the gauge indices 
$\alpha$ as equivalent.
The values of the nonvanishing elements $\Phi^{i\alpha}$ 
 and $\Psi_{\alpha i}$ of the eigenvectors in Eq.~(\ref{eigen}) 
 are explicitly determined by the constraints 
 (\ref{WZvac7}) and (\ref{WZvac8}): 
\begin{eqnarray}
& |\Phi^{i \alpha}|^2-|\Psi_{\alpha i}|^2=c,
\label{ct1}& \\
& \Psi_{\alpha i} \Phi^{i\alpha} = b~({\rm no~sum}).\label{ct2}& 
\end{eqnarray}
Let us again assume $b=0$ and $c \neq 0$.
Then, if $c>0$, we have $|\Phi^{i\alpha}|=\sqrt{c}$ 
 and $\Psi_{\alpha i}=0$, 
 and if $c<0$, we have
 $\Phi^{i\alpha}=0$ and $|\Psi_{\alpha i}|=\sqrt{-c}$.
Thus, we obtain the same result as in the previous section.  
In the case $b \neq 0$ and $c=0$, 
we have $\Phi^{i\alpha} = \sqrt be^{i\theta}$ 
 and $\Psi_{\alpha i} = \sqrt be^{-i\theta}$, with 
an arbitrary phase $\theta$.

For example, let us consider the case $M=2,~N=4$ again. 
In this case, there are ${}_4C_2=6$ solutions
 for both $\Phi$ and $\Psi$:
\begin{eqnarray}
\Phi=&\begin{pmatrix}
      \Phi^{11} & 0   \\
      0   & \Phi^{22} \\
      0   & 0   \\
      0   & 0
     \end{pmatrix},~
\begin{pmatrix}
      \Phi^{11} & 0   \\
      0   & 0   \\
      0   & \Phi^{32} \\
      0   & 0
\end{pmatrix},~
\begin{pmatrix}
      \Phi^{11} & 0   \\
      0   & 0   \\
      0   & 0   \\
      0   & \Phi^{42}
\end{pmatrix},~\nonumber \\
&\begin{pmatrix}
      0   & 0   \\
      \Phi^{21} & 0   \\
      0   & \Phi^{32}   \\
      0   & 0
\end{pmatrix},~
\begin{pmatrix}
      0 & 0   \\
      \Phi^{21}   & 0 \\
      0   & 0   \\
      0   & \Phi^{42}
\end{pmatrix},~
\begin{pmatrix}
      0   & 0 \\
      0   & 0 \\
      \Phi^{31} & 0 \\
      0   & \Phi^{42}
     \end{pmatrix}, \label{eigensol1}
\end{eqnarray}
and
\begin{eqnarray}
\Psi=
&\begin{pmatrix}
 \Psi_{11} & 0 & 0 & 0 \\
 0           & \Psi_{22} & 0 & 0
\end{pmatrix},~
\begin{pmatrix}
 \Psi_{11} & 0 & 0 & 0 \\
 0           & 0 &\Psi_{23} & 0
\end{pmatrix},~ 
\begin{pmatrix}
 \Psi_{11} & 0 & 0 & 0 \\
 0           & 0 & 0 &\Psi_{24}
\end{pmatrix},& \nonumber \\
&\begin{pmatrix}
 0 & \Psi_{12} & 0 & 0  \\
 0 & 0 & \Psi_{23} & 0 
\end{pmatrix},~
\begin{pmatrix}
 0 & \Psi_{12} & 0 & 0 \\
 0 & 0 & 0 &\Psi_{24} 
\end{pmatrix},~
\begin{pmatrix}
 0 & 0 & \Psi_{13} & 0 \\
 0 & 0 & 0 &\Psi_{24}
\end{pmatrix}.& \label{eigensol2}
\end{eqnarray}
Clearly, for $b=0$ and $c \neq 0$, we obtain
 $|\Phi^{i\alpha}|=\sqrt{c}$ and $\Psi_{\alpha i}=0$ 
 if $c>0$ and $\Phi^{i\alpha}=0$ and $|\Psi_{\alpha i}|=\sqrt{-c}$
 if $c<0$.

\section{Massive HK quotient with respect to $SU(M)$ gauge group}
In this subsection, we construct the massive HK sigma model 
 with the $SU(M)$ gauge group. We eliminate the vector multiplets 
 in the superfield formalism 
 and find that this model does not have discrete vacua.
Because we can carry out the same analysis in the case of
 Wess-Zumino gauge 
 as in the case of the $U(M)$ gauge group, 
 we do not repeat it.

\subsection{Massive HK Sigma Model with $SU$ gauge group}
In this section, we consider ${\cal N}=2$ SUSY QCD with 
 $N$ flavors and the $SU(M)$ gauge group.
We employ the same matter field content as with $T^* G_{N,M}$, 
 but here, the gauge multiplets take values in the Lie 
 algebra $SU(M)$, i.e., 
 $V= V^A T_A$ and $\Sigma = \Sigma^A T_A$, where 
 $T_A$ represents the generators of $SU(M)$.
Then, the Lagrangian is given by 
\beq
&& {\cal L} = \int d^4 \theta
 \left[ \tr (\Phi\dagg\Phi e^V )  
 + \tr (\Psi\Psi\dagg e^{-V}) \right]  
 \non &&\hs{5} 
 + \left[ \int d^2\theta \,
       \left(\tr ( \Sigma \Psi \Phi ) 
      + \sum_{a=1}^{N-1} m_a \tr (\Psi H_a \Phi)\right) 
         + {\rm c.c.}\right] .
\label{linear-SU}
\eeq
There are no FI parameters, because there is no 
 $U(1)$ gauge symmetry. 
The $SU(M)$ gauge transformation 
 is obtained in the same way as in the $U(M)$ case, 
 and it is complexified to $SU(M)^{\bf C} = SL(M,{\bf C})$.
This model has the additional $U(1)_{\rm D}$ flavor symmetry 
\beq
 \Phi \to \Phi' = e^{i \lam} \Phi \;, \hs{5}
 \Psi \to \Psi' = e^{- i \lam} \Psi , \label{U(1)D}
\eeq
which was gauged in the $U(M)$ case. 
In the massless limit, $m_a \rightarrow 0$, 
 the total flavor symmetry is 
 $U(N) = SU(N) \times U(1)_{\rm D}$. 
With a non-vanishing mass term,
 this $U(N)$ is explicitly broken down to $U(1)^N$.

We eliminate all auxiliary superfields 
 in the superfield formalism.
Then, the equations of motion for $V$ and $\Sigma$ 
 read\footnote{ 
 Here we take the Maurer-Cartan 1-form 
 $\delta X = \delta X^A T_A \equiv e^{-V} \delta e^V$ as 
 an infinitesimal parameter of variation, as in the case of
 $T^*G_{N,M}$. 
Then, the equations 
 $\delta \tr (\Phi\dagg\Phi e^V) 
 = \tr(\Phi\dagg\Phi e^V T_A) \delta X^A$ 
 and $\delta \tr (\Psi\Psi\dagg e^{-V}) = 
 - \tr(e^{-V}\Psi\Psi\dagg T_A) \delta X^A$ hold.
}
\begin{eqnarray}
{\partial {\cal L} \over \partial X^A} 
 &=& \tr [(\Phi\dagg\Phi e^V - e^{-V} \Psi\Psi\dagg)T_A ] = 0 \; , 
 \label{EOM-general-V} \\
{\partial {\cal L} \over \partial \Sigma^A} 
 &=& \tr (\Psi \Phi T_A) = 0\;, \label{EOM-general-sigma}
\end{eqnarray}
respectively. 
These equations imply
\beq
 && \Phi\dagg\Phi e^V - e^{-V} \Psi\Psi\dagg = C {\bf 1}_M 
     \; \label{EOM-V2} , \\ 
 && \Psi \Phi = B {\bf 1}_M \;, \label{EOM-sigma2} 
\eeq
 respectively, 
 where $C(x,\theta,\thb)$ and $B(x,\theta,\thb)$ are 
 vector and chiral superfields in 
 the ${\cal N}=1$ superfield formalism.\footnote{
We can understand how $C$ satisfies 
 the condition for a vector superfield, $C\dagg =C$, 
 if we rewrite Eq.~(\ref{EOM-V2}) as
 $e^{V\over 2}\Phi\dagg\Phi e^{V\over 2} 
 - e^{- {V\over 2}} \Psi\Psi\dagg e^{-{V \over 2}} 
 = C {\bf 1}_M$. 
}

The gauge field $V$ can be obtained in 
 terms of the dynamical fields from Eq.~(\ref{EOM-V2}) 
as
\beq
 e^V = {1 \over 2} (\Phi\dagg\Phi)^{-1} 
  \left(C {\bf 1}_M  \pm \sqrt {C^2 {\bf 1}_M 
     + 4 \Phi\dagg\Phi \Psi\Psi\dagg}\right) \,. 
  \label{sol-V2}
\eeq 
Then, because the equation $\det e^V =1$ holds, 
we obtain the equation 
\beq
  \det \left(C {\bf 1}_M \pm \sqrt{C^2 {\bf 1}_M  
              + 4 \Phi\dagg\Phi \Psi\Psi\dagg} \right) 
   =  2^M \det (\Phi\dagg \Phi) \;,\label{c}
\eeq
which enables us to express $C$ in terms of dynamical fields 
implicitly: $C = C(\Phi,\Phi\dagg;\Psi,\Psi\dagg)$. 
On the other hand, Eq. (\ref{EOM-sigma2}) implies
\beq
 B = \1{M} \tr (\Phi \Psi) \;. \label{b}  
\eeq

Substituting the solution (\ref{sol-V2}) 
back into the Lagrangian
(\ref{linear-SU}), we obtain the K\"ahler potential 
\begin{equation}
 K = \pm \tr \sqrt{C^2 (\Phi,\Phi\dagg;\Psi,\Psi\dagg) {\bf 1}_M 
     + 4 \Phi\dagg\Phi \Psi\Psi\dagg} 
  \;, \label{kahler-SU}
\end{equation}
 with $C$ satisfying the constraint (\ref{c}).
We should choose the plus sign
 to realize the positivity of the metric.
Using the $N\times N$ meson matrix ${\cal M} = \Phi\Psi$, 
 the \kahler potential can be rewritten as~\cite{AP}
\beq
 &&K = {M\over N} \tr_{N\times N} 
       \sqrt{C^2 {\bf 1}_N + 4{N\over M} {\cal M} {\cal M}\dagg} \;.
  \label{kahler-SU2}
\eeq
The \kahler potential (\ref{kahler-SU}) or (\ref{kahler-SU2}) 
 is strictly invariant under the full global (flavor) 
 $U(N)$ symmetry, because the mass term in the Lagrangian 
 (\ref{linear-SU}) does not affect the D-term. 

Let us stipulate the complex gauge symmetry 
 $SU(M)^{\bf C} = SL(M,{\bf C})$ 
 to express the Lagrangian
 in terms of independent superfields. 
We can employ a gauge that is similar to that in
 the $b\neq 0$ case in $T^* G_{N,M}$:
\beq
 \Phi = \sig \begin{pmatrix} {\bf 1}_M \cr \ph \end{pmatrix}P \;, \hs{5} 
 \Psi = P ({\bf 1}_M, \psi) \rho \;, \hs{5}
 P = ({\bf 1}_M + \psi\ph)^{-\1{2}} \;. \label{fixing-SU}
\eeq
Here $\ph$ and $\psi$ are 
 $[(N-M)\times M]$- and $[M\times (N-M)]$-matrix chiral 
 superfields, respectively, and
 $\sig$ and $\rho$ are chiral superfields 
 satisfying $\sig \rho = B$ from Eq.~(\ref{b}). 
We can consider $\sig$ and $\rho$ to be
 independent fields among the three fields 
 $\sigma, \rho$ and $B$. 

Substituting Eq.~(\ref{fixing-SU}) into 
 the \kahler potential (\ref{kahler-SU}), 
 we obtain the \kahler potential in terms of the independent fields 
 $\ph,\psi,\rho,\sigma$ and their conjugates. 
The superpotential also can be calculated as
\beq
 W = \sum_a m_a \sigma \rho \; 
   \tr \left[
    H_a \begin{pmatrix} {\bf 1}_M \cr \ph \end{pmatrix} ({\bf 1}_M + \psi\ph)^{-1}
    ({\bf 1}_M, \psi)  
   \right] \;. \label{superpot-SU}
\eeq
This target manifold has the isometry
 $U(N) =SU(N) \times U(1)_{\rm D}$, 
 in which the $SU(N)$ part is the same as that in the case
 of $T^* G_{N,M}$. 
The \kahler potential does not undergo 
 the \kahler transformation. 
As for the symmetry of the Lagrangian, 
 the superpotential is invariant under 
 the $U(1)$ fiber symmetry originating from (\ref{U(1)D}),
\beq
 \sig \to \sig' = e^{i \lam} \sig \;, \hs{5} 
 \rho \to \rho' = e^{- i \lam} \rho \;,
\eeq
in addition to the $U(1)^{N-1}$ symmetry of the massive 
 $T^* G_{N,M}$ model. 
Gauging this $U(1)_{\rm D}$ symmetry,
 we obtain 
 the $T^*G_{N,M}$ model again, by definition. 
Gauging the $U(1)_{\rm D}$ symmetry implies fixing 
 $B$ and $C$ in the constraints 
 (\ref{c}) and (\ref{b}) as constants.
Then, these constraints fixing
 become identical to those in the
 $T^*G_{N,M}$ case (\ref{EOM-V}) and (\ref{EOM-sig}),
 respectively.

The above discussion clarifies the bundle structure: 
 The set of $\sig$ and $\rho$ 
 constitutes a fiber of quaternion with manifold as a whole
 being the quaternionic line bundle over $T^* G_{N,M}$.

\medskip
It is interesting that we can define the same model using a 
 Lagrangian similar to that in the $T^*G_{N,M}$ case 
 if we promote the FI-parameters $b$ and $c$ in 
 the $T^*G_{N,M}$ model to the
 chiral and vector superfields 
 $B(x,\theta,\thb)$ and $C(x,\theta,\thb)$, 
 respectively,  
 constituting an ${\cal N} =2 $ vector multiplet 
 without a kinetic term. 
Then, the Lagrangian has precisely the 
 same field content as that for $T^* G_{N,M}$
\footnote{${\cal N}=2$ SUSY with this action can be understood 
 by considering the transformation of the $U(1)_J$ group 
 which is the subgroup of $SU(2)_R$ under which 
 $\theta\rightarrow e^{i\alpha}\theta,~\Phi\rightarrow 
 e^{i\alpha}\Phi,~\Psi\rightarrow e^{-i\alpha}\Psi,~V\rightarrow V,
~\Sigma\rightarrow\Sigma,~C\rightarrow C$ and $B\rightarrow B.$}
:
\beq
&& {\cal L} = \int d^4 \theta
 \left[ \tr (\Phi\dagg\Phi e^V )  
 + \tr (\Psi\Psi\dagg e^{-V}) - C\, \tr V \right]  \non
&&\hs{5} + \left[ \int d^2\theta \,
       \left(\tr \left\{ \Sigma (\Psi \Phi - B {\bf 1}_M) \right\}  
      + \sum_{a=1}^{N-1} m_a \tr (\Psi H_a \Phi)\right) 
         + {\rm c.c.}\right] .
\label{linear-SU2}
\eeq
In this case, the equations of motion for 
$V$, $\Sig$, $B$ and $C$ read
\begin{eqnarray}
 && {\partial {\cal L} \over \partial V} 
 = \Phi\dagg\Phi e^V - e^{-V} \Psi\Psi\dagg - C {\bf 1}_M = 0 \; 
 , \hs{5}
  {\partial {\cal L} \over \partial \Sigma} 
 = \Psi \Phi - B {\bf 1}_M = 0\;, \\
 && {\del {\cal L} \over \del B} =  - \tr \Sigma = 0 \;, \hs{5}
  {\del {\cal L} \over \del C} =  - \tr V = 0 \;. 
\end{eqnarray}
We thus have the same constraints as in 
 the model (\ref{linear-SU}). 
The technique of 
 promoting an FI-parameter to a superfield was
 used in Ref. \citen{HKN} to construct 
 the complex canonical line bundle over $G_{N,M}$ 
 in an ${\cal N}=1$ (four SUSY) theory, where 
 an additional superfield $\sigma$ relative to 
 $G_{N,M}$ appeared and 
 was identified with the fiber of a complex line.
Returning to ${\cal N}=2$, we obtain the total manifold as the
 cotangent bundle over the complex line bundle on $G_{N,M}$,
 where $\rho~(\psi)$ serves as the cotangent fiber over 
 $\sigma~(\varphi)$.
We have thus found that the space possesses two equivalent bundle
 structures.
We may need to transform $\sigma$ and $\rho$ to 
 some fields $\sigma'$ and $\rho'$ in order
 to avoid coordinate singularities, as 
 in the four SUSY case~\cite{HKN}.

\medskip
\underline{Other gauge groups: $SO$ and $Sp$}

The equations of motion for $V$ and $\Sigma$ takes the same 
 form as Eqs.~(\ref{EOM-general-V}) and 
 (\ref{EOM-general-sigma}) for {\it any} gauge group. 
However, the right-hand side of 
 Eq.(\ref{EOM-V2}) must be a 
symmetric tensor for the 
 $SO(M)$ gauge group, and the right-hand side of 
 Eq.(\ref{EOM-sigma2}) must be a pseudo-symmetric tensor  
 for the $Sp(M)$ gauge group.

\subsection{Vacua of $SU$ gauge theories}

We now seek the vacua of the HK sigma model with the $SU$ 
 gauge group.
The superpotential (\ref{superpot-SU}) of this model 
 can be rewritten as
\beq
 W = \sigma \rho \sum_{a=1}^{N-1} \sum_{n=0}^{\infty} (-1)^n  
    m_a \tr \left[ H_a 
     \begin{pmatrix} (\psi\ph)^n & 0 \cr
                         0 & (\ph \psi)^{n+1} \end{pmatrix}
      \right]
  \equiv \sigma \rho W_U \;,
\eeq
where $W_U$ (times $b$) denotes the superpotential 
 (\ref{superpot2}) or (\ref{superpot3}) of 
 the $U(M)$ gauge group with $b \neq 0$.
The derivatives of the superpotential with respect to the fields 
 are given by
 $\del_{\psi} W = \sig \rho \del_{\psi} W_U$, 
 $\del_{\ph} W = \sig \rho \del_{\ph} W_U$, 
 $\del_{\rho} W = \sig W_U$ and $\del_{\sig} W = \rho W_U$. 
The vacuum condition is given by $\sig = \rho =0$, 
 because $\del W_U = 0$ holds only at $\ph=\psi^T=0$,
 as found in the previous section, 
 but we have $W_U \neq 0$ there.
Therefore, this model has no discrete vacua, 
 and so there cannot be any wall solution.

Vacua of the massive HK sigma model with the $SU(M)$ gauge group 
 may have a similarity with that with the $U(M)$ gauge group, 
 the massive $T^*G_{N,M}$ model, in a particular limit.
In the latter model, there exist the FI-parameters $b$ and $c$, 
 which represent the radius of the base $G_{N,M}$, 
 and the discrete vacua are attached to this manifold. 
In the limit of vanishing $b$ and $c$, 
 a singularity appears. 
All of the discrete vacua go to that point, 
 and as a result there are no sets of discrete vacua.
Because the massive HK sigma model with the $SU(M)$ gauge group 
 cannot contain the FI parameters from the beginning, 
 we believe that this model has no discrete vacua.
It is thus seen that the FI-parameters seem to play an 
 essential role in determining the existence of discrete vacua.

%
%
\section{Formulation in the harmonic superspace}
In this section we formulate our models with 
 $U(M)$ and $SU(M)$ gauge symmetries 
 using the HSF.
Massless nonlinear sigma models with toric and non-toric 
 HK manifolds
 using the HSF were originally considered in 
 Refs.~\citen{ivanov-EH} and \citen{ivanov-TN}
 using the quotient construction.
Massive nonlinear sigma models with toric HK manifolds, 
 like the Taub-NUT \cite{ketov} metric and 
 the Eguchi-Hanson metric, have also been 
 studied \cite{ketov, ANNS}. 
We construct massive HK sigma models with 
 our non-toric HK manifolds using HSF.
One advantage of constructing massive nonlinear sigma models 
 using the HSF is that ${\cal N}=2$ SUSY is manifest
 in this case.
It is easy to extend these models to cases with other flavor
 and gauge symmetries.

We first present the actions of the $U(M)$ and $SU(M)$ models 
 in terms of the harmonic superfields.
Next, we derive their component actions in the Wess-Zumino gauge.
We find that the scalar potential is represented 
 by auxiliary fields
 in the hypermultiplet analytic superfield 
 and we obtain a vacuum condition that 
 is identical to that in 
 the ${\cal N}=1$ formulation.
Finally, we obtain the same result regarding the vacua
 as in the ${\cal N}=1$ formulation.
We follow the notation of Ref.~\citen{ANNS}.

\subsection{Massive HK sigma model with 
 $U(M)$ and $SU(M)$ gauge groups}

First of all, we consider the model with $U(M)$ gauge 
 symmetry.
The HSF action for the sigma models with $U(M)$ and $SU(M)$ 
 gauge symmetries 
 can be described by two kinds of analytic 
 superfields, which are hypermultiplet and vector multiplet analytic 
 superfields.
A hypermultiplet is defined by the analytic superfield
 $q^{+a\alpha}~(a=1,\cdots,N,~\alpha=1,\dots,M)$ 
 in the fundamental representation
 of the $SU(N)$ flavor symmetry\footnote{
 To avoid confusion with the $SU(2)_R$ indices $i=1, 2$ of 
 the harmonic variable, we denote the $SU(N)$ flavor indices by 
 $a=1, \cdots, N$ in the following. 
 } and the $U(M)$ gauge symmetry,
 which is a function in the harmonic analytic 
 ${\cal N}=2$ superspace
\begin{eqnarray}
\zeta_A=(x_A^\mu,\theta^+,\bar{\theta}^+,u_i^{\pm})\,, \label{hss}
\end{eqnarray}
where the coordinates 
 $u^{+i},u^{-i}, u^{+i}u_{i}^-=1~(i=1,2)$ are the
 $SU(2)_R/U(1)_r$ harmonic variables.
Here $U(1)_r$ is a diagonal subgroup of $SU(2)_R$.
The $U(M)$ gauge transformation for the hypermultiplet is given by
\begin{eqnarray}
q^{+a\alpha} \rightarrow 
 (e^{-i \lambda(\zeta_A,u)})^\alpha_{~\beta} q^{+a\beta}\,,
 \label{eq:gaugetrans1}
\end{eqnarray}
where $\lambda=\lambda^A T_A $ [$T_A$ being the generator of
 the $U(M)$ gauge symmetry] is the real analytic superfield
 with $U(1)_r$ charge $0$, which
 represents the gauge transformation parameter.
The vector multiplet $V^{++}=V^{A++}T_A$ is defined as the real
 analytic superfield, that is, $\widetilde{V^{++}}=V^{++}$, where
 the tilde denotes conjugation which is 
 the combination of complex conjugation 
 and the star conjugation \cite{Ivanov2, ANNS}.
The action of the tilde conjugation 
 on $\zeta_A$ is defined as 
\begin{eqnarray}
\widetilde{x_A^\mu}=x_A^\mu,~\widetilde{\theta^+}=\bar{\theta}^+,
~\widetilde{\bar{\theta}^+}=-\theta^+,~\widetilde{u_i^{\pm}}=u^{\pm i}\,,
~\widetilde{u^{\pm i}}=u^{\pm}_i\,.
\end{eqnarray}
The vector multiplet superfield transforms under the $U(M)$ gauge 
 transformation as 
\begin{eqnarray}
 V^{++}\rightarrow e^{-i\lambda}V^{++}e^{i\lambda}
 -i e^{-i\lambda}D^{++}e^{i\lambda}\,,
 \label{eq:gaugetrans2}
\end{eqnarray}
where $D^{++}$ is the covariant derivative defined by
\begin{eqnarray}
 D^{++}=\partial^{++}-2 i \theta^+ \sigma^\mu {\bar{\theta}}^+ 
        \partial_\mu^A,~~~~~ 
 \partial^{++}=u_i^+\frac{\partial}{\partial u_i^-}\,,~~~
 \partial_\mu^A=\frac{\partial}{\partial x_A^\mu}\,. 
        \label{covdhss}
\end{eqnarray}
The infinitesimal forms of (\ref{eq:gaugetrans1}) 
and (\ref{eq:gaugetrans2}) are 
\begin{eqnarray}
 \delta q^+&=&-i\lambda q^{+}\,,\\
 \delta V^{++}&=&D^{++}\lambda-i[\lambda,V^{++}]\,.
\end{eqnarray}
The action of the massive $T^*G_{N,M}$ model is invariant under
 these transformations, and it is given by
\begin{eqnarray}
 S &=& -\displaystyle\int d\zeta_A^{(-4)} du   
        {\Bigg \{}{\widetilde{q^+}}_{a\alpha}
           (D^{++}+i V^{++})^\alpha_{~\beta}q^{+a\beta}
       +\xi^{++} {\rm tr} (V^{++}) \nonumber \\ 
   & & ~~~~~~~~
       -\sum_{A=1}^{N-1}{\widetilde{q^+}}_{a\alpha}(\theta^{+2}{\bar{m}}_A
       -{\bar{\theta}}^{+2}m_A)(H_A)^a_{~b} q^{+b\alpha}           
       {\Bigg \}}\,, \label{TSG}
\end{eqnarray}
where 
 $d\zeta_A^{(-4)}du=d^4 x_A d^2 \theta^{+} d^2 \bar{\theta}^{+}du$
 is the measure of the integration over analytic 
 superspace given (\ref{hss}), 
 $\xi^{++}=\xi^{(ij)}u_{(i}^+u_{j)}^+$ 
 is the coefficient of the FI term 
 [which is the $SU(2)_R$ triplet],
 $m_A$ represents complex mass parameters, and
 $H_A$ represents the diagonal generators of the Cartan 
 subalgebra of $SU(N)$ defined in (\ref{linear}).
Here, the tilde conjugation acts not on the gauge and flavor
 indices but on the $U(1)_r$ charge.
The quantity $V^{++}$ serves as the Lagrange multiplier.
Integrating the vector multiplet, it gives a constraint for the
 hypermultiplet.
At the component level, it gives a constraint on the scalars
 and fermions, and it makes the target manifold nontrivial, 
 as we see in the next subsection.
We have absorbed a common hypermultiplet mass
 through a shift of the analytic superfield $V^{++}$.
In the limit $m_A \rightarrow 0$,
 the action (\ref{TSG}) becomes 
 that of the massless $T^*G_{N,M}$
 model, whose isometry is $SU(N)$. 
The mass term explicitly 
 breaks $SU(N)$ down to $U(1)^{N-1}$.
This feature is identical to that in 
 the case of the ${\cal N}=1$ formalism. 

Next we consider the model with $SU(M)$ gauge symmetry.
This quotient action can be easily obtained by restricting
 the gauge symmetry $U(M)$ to $SU(M)$ in the action 
 (\ref{TSG}).
In this case, 
 the FI term does not exist,
 since the theory no longer possesses $U(1)$ gauge symmetry.
The action is given by
\begin{eqnarray}
 S &=& -\displaystyle\int d\zeta_A^{(-4)} du   
        \{{\widetilde{q^+}}_{a\alpha}
           (D^{++}+i V^{++})^\alpha_{~\beta}q^{+a\beta} \nonumber \\ 
   & & ~~~~~~~~
       -\sum_{A=1}^{N-1}{\widetilde{q^+}}_{a\alpha}(\theta^{+2}{\bar{m}}_A
       -{\bar{\theta}}^{+2}m_A)(H_A)^a_{~b}q^{+b\alpha}           
       \}\,. \label{TSG2}
\end{eqnarray}
Note again that the theory has an additional $U(1)$ global 
 symmetry, like the ${\cal N}=1$ formulation. 
The total global symmetry in the $m_A=0$ case is $U(N)$.
The global symmetry $U(N)$ is broken down to $U(1)^N$ 
 when there exists a non-vanishing mass term. 

We can obtain the nonlinear sigma model action in 
 terms of the independent harmonic superfields by fixing the 
 gauge symmetry and solving the constraint, as in the 
 ${\cal N}=1$ formulation.
The resulting action can be described by the independent analytic
 superfields of hypermultiplets.
However, it is often difficult to decompose the action 
 consisting independent superfields into
 a component action, because the kinematical part of the 
 equations of motion, which is necessary to obtain the component
 action, is difficult to solve. 
Instead, we can easily solve the kinematical part of the equations
 of motion when we employ the Wess-Zumino gauge and
 can obtain the explicit form of the component action.
This gauge is particularly convenient to understand 
the vacua of theories in the HSF.

\subsection{Scalar potential in the Wess-Zumino gauge}
\newcommand{\owm}{\bar{\widetilde{m}}}
\newcommand{\wm}{\widetilde{m}}

We now focus on the $U(M)$ case. 
In the following, we write the mass matrix as 
 $\Sigma_{A=1}^{N-1}m_A H_A\equiv\widetilde{m}$ and  
 $\Sigma_{A=1}^{N-1}\bar m_A H_A\equiv\bar {\widetilde{m}}$\,. 
In this case,
 the action can be rewritten as 
\begin{eqnarray}
S=\displaystyle\int d\zeta_A^{(-4)}du{\Bigg \{}
 \bar{q}^+_{a\alpha}(D_{cc}^{++}+iV^{++})^\alpha_{~\beta}
 q^{+a\beta}-\xi^{++} \tr (V^{++}){\Bigg \}}\,, \label{grhssaction}
\end{eqnarray}
where we have used $\widetilde{q^+}=-\bar{q}^+$ and
 the definition of the covariant derivative including
 the mass term 
\begin{eqnarray}
D_{cc}^{++}\equiv \partial^{++}
 -2i\theta^+\sigma^\mu\bar{\theta}^+\partial_\mu^A
 -(\theta^{+2}\bar{\widetilde{m}}-\bar{\theta}^{+2}\widetilde{m})\,.
\end{eqnarray}
The equations of motion are
\begin{eqnarray}
0&=&(D_{cc}^{++}+iV^{++})^\alpha_{~\beta}q^{+a\beta}, \label{eqm1}\\
0&=&\bar{q}^+_{a\alpha}i(T_A)^\alpha_{~\beta}q^{+a\beta} 
 -\xi^{++}{\rm tr}(T_A)\,,\label{eqm2}
\end{eqnarray}
where (\ref{eqm1}) includes kinematical and dynamical parts of 
 the equations of motion, and (\ref{eqm2}) is a constraint.
In the HK sigma model, we are only interested in the 
 bosonic components.
To obtain the bosonic Lagrangian, we have to solve the
 kinematical part of the equations of motion to eliminate the infinite
set of auxiliary fields.
To solve the kinematical part, we substitute the Grassmann 
 expansion of the analytic superfields into the above equations.
In the Wess-Zumino gauge, the bosonic components in the
 Grassmann expansion of the analytic superfields 
 $q^{+a\alpha}$ and $V^{++}$ are given by
\begin{eqnarray}
q^{+a\alpha}&=&F^{+a\alpha}+\theta^{+2}M^{-a\alpha}
 +\bar{\theta}^{+2}N^{-a\alpha}
 +i\theta^+\sigma^\mu\bar{\theta}^+A_\mu^{-a\alpha}
 +\theta^{+2}\bar{\theta}^{+2}D^{(-3)a\alpha}, \label{exp1} \\
V^{++}&=&\theta^{+2}\bar{M}_v+\bar{\theta}^{+2}M_v
  +i\theta^+\sigma^\mu\bar{\theta}^+(-2V_\mu)
  +\theta^{+2}\bar{\theta}^{+2}D_v^{(--)},\label{exp2}
\end{eqnarray}
where each component in $V^{++}$ is a Lie algebra-valued field, for
 example $M_v=M_v^A T_A$, and 
 $D_v^{(--)}\equiv D_v^{(ij)}u_i^-u_j^-$.
Note that the components in $V^{++}$ do not depend on 
 the harmonic variables,
 whereas the components in $q^+$ include the harmonic variables.
Substituting them into (\ref{eqm1}), we obtain the 
 following kinematical 
 part of the equations of motion
\begin{eqnarray}
0&=&\partial^{++}F^{+a\alpha}\,,\label{eq1} \\
0&=&\partial^{++}M^{-a\alpha}-\owm^a_{~b}F^{+b\alpha}
 +i\bar{M_v}^\alpha_{~\beta}F^{+a\beta}\,,\label{eq2} \\
0&=&\partial^{++}N^{-a\alpha}+\wm^a_{~b}F^{+b\alpha}
 +i{M_v}^\alpha_{~\beta}F^{+a\beta}\,,\label{eq3} \\
0&=&\partial^{++}A_\mu^{-a\alpha}-2\partial_\mu^A F^{+a\alpha}
 -2i{V_\mu}^\alpha_{~\beta}F^{+a\beta}\,.\label{eq4}
\end{eqnarray}
The equations (\ref{eq1})--(\ref{eq4}) can be easily solved,
 and we find
\begin{eqnarray}
F^{+a\alpha}(x_A,u)&=&f^{ia\alpha}(x_A)u_i^+\,, \label{eq6} \\
M^{-a\alpha}(x_A,u)&=&\owm^a_{~b}f^{ib\alpha}(x_A)u_i^-
 -i({\bar{M}}_v)^\alpha_{~\beta}f^{ia\beta} (x_A)u_i^-\,, \label{eq7} \\
N^{-a\alpha}(x_A,u)&=&-\wm^a_{~b}f^{ib\alpha}(x_A)u_i^-
 -i(M_v)^\alpha_{~\beta}f^{ia\beta}(x_A)u_i^-\,, \label{eq8} \\
A_\mu^{-a\alpha}(x_A,u)&=&2\partial_\mu^A f^{ia\alpha}(x_A)u_i^-
 +2i(V_\mu)^\alpha_{~\beta} f^{ia\beta}(x_A)u_i^-\,.\label{eq9}
\end{eqnarray} 
Note that the Lagrange multipliers $V_\mu$ and $M_v$ remain, 
 although
 an infinite set of auxiliary fields have been eliminated.

At this stage, we can write down the component action.
Substituting 
 the Grassmann expansions (\ref{exp1}) and (\ref{exp2})
 into the action (\ref{grhssaction}),
 using the equations of motion (\ref{eq1})--(\ref{eq4}), and
 integrating the Grassmann variable,
 the action (\ref{grhssaction}) can be put into the 
 following form:
\begin{eqnarray}
S&=&\displaystyle\int d^4 x_A du{\Bigg \{}
 -\bar{F}^+_{a\alpha}\partial^\mu_A A_\mu^{-a\alpha}
 -i\bar{F}^+_{a\alpha}(V^\mu)^\alpha_{~\beta}A^{-a\beta}_\mu 
 -\bar{F}^+_{a\alpha}\owm^a_{~b}N^{-b\alpha} \nonumber \\
&&
 +i\bar{F}^{+}_{a\alpha}(\bar{M}_v)^\alpha_{~\beta}N^{-a\beta}
 +\bar{F}^+_{a\alpha}\wm^a_{~b}M^{-a\alpha}
 +i\bar{F}^+_{a\alpha}(M_v)^\alpha_{~\beta}M^{-a\beta} \nonumber \\
&&
 +D_v^{A(--)}(i\bar{F}^+_{a\alpha}(T_A)^\alpha_{~\beta}F^{+a\beta}
 -\xi^{++}{\rm tr}(T_A)){\Bigg \}}\,. \label{grhssaction2}
\end{eqnarray}
Then, using the conjugates of the
 equations of motion (\ref{eq2})-(\ref{eq4}),
\begin{eqnarray}
\bar{M}^+_{a\alpha}&=&\bar{M}^i_{~a\alpha}u_i^+
 =\bar{F}^+_{b\alpha}\wm^b_{~a}+i\bar{F}^+_{~a\beta}(M_v)^\beta_{~\alpha}\,, \\
\bar{N}^+_{a\alpha}&=&\bar{N}^i_{~a\alpha}u_i^+
 =-\bar{F}^+_{b\alpha}\owm^b_{~a}
 +i\bar{F}^+_{~a\beta}(\bar{M}_v)^\beta_{~\alpha}\,, \\
\bar{A}_{\mu a\alpha}^+&=&\bar{A}^i_{\mu a\alpha}u_i^+
 =2(\partial_\mu^A\bar{F}^+_{a\alpha}
 -i\bar{F}^+_{a\beta}(V_\mu)^\beta_{~\alpha})\,,
\end{eqnarray}
where $\bar{F}^+=\bar{f}^iu_i^+$,
we can rewrite the action in the simple form
\begin{eqnarray}
S&=&\displaystyle\int d^4x_A du {\Bigg \{}
 \bar{A}^{\mu+}_{a\alpha}A_\mu^{-a\alpha}
 +\bar{M}^+_{a\alpha}M^{-a\alpha}+\bar{N}^+_{a\alpha}N^{-a\alpha} \nonumber \\
&& +D_v^{A(--)}(i\bar{F}^+_{a\alpha}(T_A)^\alpha_{~\beta}F^{+a\beta}
 -\xi^{(++)}{\rm tr}(T_A)){\Bigg \}}\,.
\end{eqnarray}
The harmonic variables can be integrated easily by using the 
 following formulas:
\begin{eqnarray}
\displaystyle \int du u_i^+ u_j^-&=&\frac{1}{2}\epsilon_{ij},\\
\displaystyle \int du u_i^+u_j^+u_k^-u_l^-
 &=&\frac{1}{6}(\epsilon_{il}\epsilon_{jk}+\epsilon_{ik}\epsilon_{jl})\,.
\end{eqnarray}
The action after integration over $u$ is
\begin{eqnarray}
S&=&\displaystyle \int d^4x_A\left({\cal L}_{\rm kin}
 +{\cal L}_{\rm constr}
 +{\cal L}_{\rm pot}\right)\,, \\
&&{\cal L}_{\rm kin}=-\frac{1}{2}A_\mu^{ia\alpha}\bar{A}^\mu_{ia\alpha}\,,
 \label{Lk} \\
&&{\cal L}_{\rm constr}=\frac{1}{3}D^A_{v(ij)}
 \left(i\bar{f}^{(i}_{a\alpha}(T_A)^\alpha_{~\beta}f^{j)a\beta}
 -\xi^{(ij)}{\rm tr}(T_A) \right)\,,\label{Lc} \\
&&{\cal L}_{\rm pot}=-\frac{1}{2}\left(M^{ia\alpha}\bar{M}_{ia\alpha}
 +N^{ia\alpha}\bar{N}_{ia\alpha}\right)=-V\,. \label{V1}
\end{eqnarray}
After substituting (\ref{eq6})--(\ref{eq9}) 
 into (\ref{Lk})--(\ref{V1})
 and integrating over the auxiliary field $D_{v(ij)}^A$, 
 we obtain the bosonic component
 Lagrangian, 
\begin{eqnarray}
{\cal L}_{\rm kin}&=&-2(\partial_\mu^A f^{ia\alpha}
 \partial_A^\mu \bar{f}_{ia\alpha}
 -i\bar{f}_{ia\alpha}(V^\mu)^\alpha_{~\beta}\partial_\mu^Af^{ia\beta} 
 \nonumber \\
 &&+i\partial_A^\mu\bar{f}_{ia\alpha}(V_\mu)^\alpha_{~\beta}f^{ia\beta}
 +\bar{f}_{ia\alpha}(V^\mu)^\alpha_{~\beta}
 (V_\mu)^\beta_{~\gamma}f^{ia\gamma})\,, \\
V&=&\frac{1}{2}\bar{f}_{ia\alpha}\{\wm,\owm\}^a_{~b}f^{ib\alpha}
 +\frac{1}{2}\bar{f}_{ia\alpha}\{M_v,\bar{M}_v\}^\alpha_{~\beta}f^{ia\beta} 
 \nonumber \\
 &&-i\bar{f}_{ia\alpha}(\bar{M}_v)^\alpha_{~\beta}\wm^a_{~b}f^{ib\beta}
 +i\bar{f}_{ia\alpha}\owm^a_{~b}(M_v)^\alpha_{~\beta}f^{ia\beta}\,,
\end{eqnarray}
with the constraint
\begin{eqnarray}
 i\bar{f}^{(i}_{a\alpha}(T_A)^\alpha_{~\beta}f^{j)a\beta}
 -\xi^{(ij)}{\rm tr}(T_A)=0\,, \label{chss}
\end{eqnarray}
which makes the target space nontrivial.
This constraint can also be obtained as the 
 lowest-order component of (\ref{eqm2}).
Let us next eliminate the Lagrange multipliers $V_\mu$ and $M_v$.
In order to do so, we define
\begin{eqnarray}
&f^{ia\alpha}\bar{f}_{ia\beta}\equiv G^A (T_A)^\alpha_{~\beta},~~~~
 if^{ia\alpha}\dellr_\mu\bar{f}_{ia\beta}\equiv 
 H^A_\mu (T_A)^\alpha_{~\beta}\,,& \\
&\tilde{G}_{AB} \equiv \frac{1}{2}{\rm tr}
 \left(\{T_A,T_B\}T_C\right)G^C\,,& \\
&\bar{f}_{ia\beta}\wm^a_{~b}f^{ib\alpha}
 \equiv L^A (T_A)^{\alpha}{}_{\beta}\,. &
\end{eqnarray}
Using these quantities, we can rewrite the Lagrangian 
 as
\begin{eqnarray}
 {\cal L}_{\rm kin}&=&-2(\partial_\mu^A f^{ia\alpha}
 \partial_A^\mu \bar{f}_{ia\alpha}+V_\mu^AH_A^{\mu}+V_\mu^A V^{B\mu}
 \tilde{G}_{AB})\,, \label{L3}\\
V&=&\bar{f}_{ia\alpha}|\wm|^{2a}_{~~b}f^{ib\alpha}
 +M_v^A\bar{M}_v^B \tilde{G}_{AB}
 +iM_v^A \bar{L}_A-i\bar{M}_v^A L_A\,. \label{L4}
\end{eqnarray}
The equations of motion for $V_\mu$ and $M_v$ are given by
\begin{eqnarray}
\frac{\partial {\cal L}}{\partial V_\mu^A}&=&
 -H_A^{\mu}-2V^{B\mu}\tilde{G}_{AB}=0\,, \\
\frac{\partial {\cal L}}{\partial M_v^A}&=&
 \bar{M}_v^B \tilde{G}_{AB}+i\bar{L}_A=0\,.
\end{eqnarray}
Solving these and substituting into 
 (\ref{L3}) and (\ref{L4}), we finally obtain
\begin{eqnarray}
{\cal L}_{\rm kin}&=&-2\left(\partial_\mu^A f^{ia\alpha}
 \partial_A^\mu \bar{f}_{ia\alpha}
 -\frac{1}{4} H_\mu^A (\tilde{G}^{-1})_{AB} H^{B\mu}\right)\,, \\
V&=&\bar{f}_{ia\alpha}|\wm|^{2a}_{~~b}f^{ib\alpha}
 -L^A(\tilde{G}^{-1})_{AB}\bar{L}^B\,.
\end{eqnarray}

The bosonic Lagrangian of the $SU(M)$ model has the same form as
 that in the $U(M)$ model if we regard the parameters $b$ and $c$ 
 as arbitrary complex and real parameters, respectively.

\subsection{Vacua in the massive $T^*G_{N,M}$ model}
We now study vacua based on the potential (\ref{V1})
 and the constraint. 
The SUSY vacuum conditions are that the auxiliary 
 fields vanish:
\begin{eqnarray}
M^{ia\alpha}=N^{ia\alpha}=0\,. \label{co1}
\end{eqnarray}
These conditions imply vanishing values for the potential 
 (\ref{V1}) which is clearly positive definite. 
We should, of course, simultaneously 
 impose the constraint (\ref{chss}). 
It is convenient to rewrite these conditions as
\begin{eqnarray}
&f^{ia\alpha}=\left(\begin{array}{c}
	              \phi^{a\alpha} \\
                      -i\bar{\chi}^{a\alpha}  
		   \end{array}\right)\,,~~
\bar{f}^{i}_{~a\alpha}=\epsilon^{ij}\overline{f^{ja\alpha}}
                  =\epsilon^{ij}\bar{f}_{ja\alpha}
                  =\left(
                   \begin{array}{c}
		      i\chi_{a\alpha} \\
                      -\bar{\phi}_{a\alpha}
		   \end{array} 
                   \right)\,, & \\
&\xi^{(11)}=-ib^*\,,~~\xi^{(12)}=\xi^{(21)}=\frac{ic}{2},~~\xi^{(22)}=ib\,,& \\
&iM_v=\sigma\,.&
\end{eqnarray}
In terms of these variables, Eq.~(\ref{co1}) becomes 
\begin{eqnarray}
&&0=\owm^a_{~b}\phi^{b\alpha}+\bar{\sigma}^{\alpha}_{~\beta}\phi^{a\beta}\,, \label{eh1}\\
&&0=\owm^a_{~b}\bar\chi^{b\alpha}
+\bar{\sigma}^{\alpha}_{~\beta}\bar\chi^{a\beta}\,, \label{eh2}\\
&&0=
\wm^a_{~b}\phi^{b\alpha}
+\sigma^\alpha_{~\beta}{\phi}^{a\beta}\,,  
\label{eh3}\\
&&0=\wm^a_{~b}\bar\chi^{b\alpha }
+\sigma^\alpha_{~\beta}\bar\chi^{a\beta}\,.         
\label{eh4}
\end{eqnarray}
Also, the constraints in Eq.~(\ref{chss}) become 
\begin{eqnarray}
&&0=\chi_{\alpha a}\phi^{a\beta}=\bar{\phi}_{\alpha a}\bar{\chi}^{a\beta}\,, \label{eh5}\\
&&0=\chi_{\alpha a}\bar{\chi}^{a\beta}-\bar{\phi}_{\alpha a}\phi^{a\beta}    \label{eh6}
  -c\delta_\alpha^{~\beta}\,.
\end{eqnarray}

Let us solve Eqs (\ref{eh1})--(\ref{eh6}).
Using $U(M)$ rotation, $\sigma$ can be diagonalized as
\begin{eqnarray}
&\sigma={\rm diag.}(\sigma_1,\sigma,\cdots,\sigma_M)\,,& \\
&\displaystyle\sum_{\alpha=1}^M\sigma_\alpha=M\sigma^0\,.~~(\sigma=\sigma^A T_A)\nonumber&
\end{eqnarray}
Equations (\ref{eh1})--(\ref{eh4}) can be rewritten 
 in terms of the diagonal masses
 $\tilde{m}={\rm diag.}(m_1,\dots,m_N)$
 and $\sigma$ as
\begin{eqnarray}
&&0=(\bar{m}_a {\bf 1}_M+\bar{\sigma}_\alpha {\bf 1}_N)
\phi^{a\alpha}\,, \label{eh7}\\
&&0=(\bar{m}_a {\bf 1}_M+\bar{\sigma}_\alpha {\bf 1}_N)\bar{\chi}^{a\alpha}\,, \label{eh8}\\
&&0=(m_a {\bf 1}_M+\sigma_\alpha {\bf 1}_N){\phi}^{a\alpha }\,, 
\label{eh9}\\
&&0=(m_a {\bf 1}_M+\sigma_\alpha {\bf 1}_N)\bar\chi^{a\alpha }\,. 
\label{eh10}
\end{eqnarray}
Because $M^{ia\alpha}$ and $N^{ia\alpha}$ are not 
 complex conjugates of each other, 
 Eqs.~(\ref{eh7}) and (\ref{eh8}) are not 
 complex conjugates of Eqs.~(\ref{eh9}) and (\ref{eh10}). 
However, these four conditions are consistent.

Because the SUSY conditions (\ref{eh5})--(\ref{eh8}), 
together with the constraints (\ref{eh9}) and (\ref{eh10}),
 are the same form as (\ref{WZvac5})--(\ref{WZvac8}),
 we can repeat the analysis given for the
 ${\cal N}=1$ formulation in $\S$\ref{vacuasection}
 to obtain the solutions.
We thus find that there are ${}_NC_M$ solutions of SUSY vacua.

\section{Summary and discussion}

We have constructed 
 the massive $T^*G_{N,M}$ model and its generalization, 
 which are the Higgs branch of ${\cal N}=2$ SUSY gauge 
 theories with $U(M)$ and $SU(M)$ gauge groups, 
 respectively. 
The vacuum structure for the massive $T^*G_{N,M}$ model 
 has been clarified, and it was found
 to be far richer 
 than the massive $T^*{\bf C}P^{N-1}$ model.
It has discrete $N!/[M!(N-M)!]$ vacua, which are the origins of 
 the standard coordinate patches for $G_{N,M}$ and are
 represented by mutually orthogonal $M$-dimensional complex 
 planes in ${\bf C}^N$ 
 in a coordinate-independent way. 
On the other hand, the massive HK sigma model with the 
 $SU(M)$ quotient has no discrete vacua. 

Since we have discrete vacua for the massive 
 $T^*G_{N,M}$ model, 
 we believe that this model has
 domain wall configurations that are richer than 
 the massive $T^*{\bf C}P^{N-1}$ model. 
First, even for a single wall, 
 we can expect several types of solutions. 
For example, the cotangent bundle over 
 the Klein quadric, $T^*G_{4,2}$, has the  
 six vacua given in Eq.~(\ref{vac-Klein}). 
This model is expected to admit two kinds of walls. 
One is a wall connecting two vacua without any common axis, e.g., 
 the first and the sixth vacua. 
The other is a wall connecting two vacua with a common axis, e.g., 
 the first and the second vacua. 
For general massive $T^*G_{N,M}$ models, 
 we conjecture a greater variety of wall solutions. 
Second, if we have a parallel wall configuration, as
 in the massive $T^* {\bf C}P^{N-1}$ model, 
 such a configuration has several zero modes 
 corresponding to positions of the walls 
 \cite{GTT2, EFNS}. 
The number of zero modes was calculated to be $N$ 
 for the massive $T^* {\bf C}P^{N-1}$ model 
 in Ref. \citen{Le} using the index theorem. 
It is interesting to apply the approach used there to 
 the massive $T^*G_{N,M}$ model.

The second homotopy group for $T^*G_{N,M}$ is nontrivial:
 $\pi_2 (T^* G_{N,M}) \simeq \pi_2 (G_{N,M}) 
 \simeq {\bf Z}$~\cite{Pe}.
Therefore, this model is believed to admit multi-string 
 (vortex or lump) solutions, 
 a configuration of a string ending on a wall, as 
 was found in Ref. \citen{GPTT} 
 for $T^*{\bf C}P^1$, and other interesting phenomena.

Reducing the $T^*{\bf C}P^{N-1}$ model 
 to three-dimensional space-time,
 an interesting 
 mirror symmetry has been found \cite{To2}. 
Determining the non-Abelian generalization of 
 this mirror symmetry 
 using our massive $T^*G_{N,M}$ model would be
 an interesting task.

Coupling the model to supergravity is possible, and
 in that case, the target manifold is 
 the quaternionic generalization of $T^* G_{N,M}$. 
Moreover, our model can be promoted to 
 five-dimensional supergravity
 in the manner considered in Refs.~\citen{AFNS}
 and \citen{EFNS}.
This is interesting for the brane world scenario.



\vspace{1.0cm}

\noindent{\Large \bf Acknowledgements}\\

\noindent 
We thank Masashi Naganuma for collaboration 
 in the early stages of this work. 
This work is supported in part by a Grant-in-Aid 
 for Scientific Research from the Japan Ministry 
 of Education, Culture, Sports, Science and Technology 
13640269 (N.S.). 
The work of M.~N. is supported by the U.~S. Department
 of Energy under grant DE-FG02-91ER40681 (Task B).


\end{document}